\documentclass[manuscript,screen] {acmart}
\setcopyright {none}
\AtBeginDocument{%
  \providecommand\BibTeX{{%
    \normalfont B\kern-0.5em{\scshape i\kern-0.25em b}\kern-0.8em\TeX}}}

\usepackage{enumerate}
\begin{document}

\title{Generative AI's aggregated knowledge versus web-based curated knowledge}


\author{Ted Selker} 
\authornote{Both authors contributed equally to this research.}
\email{Tedselker@colorado.edu}
\orcid{0000-0002-1060-7135}
\affiliation{
 \institution{University of Colorado}
 \city{Boulder}
\state{CO}
\country{USA}
 }
 
 \author{Yunzi Wu}
 \authornotemark[1]
\email{yuwu22@lehigh.edu}
\orcid{0009-0006-4775-3693}
\affiliation{
 \institution{Lehigh University}
\city{West Bethlehem}
\state{Pennsylvania}
\country{USA}
 }



\begin{abstract}
This paper explores what kinds of questions are best served by the way generative AI (GenAI) using Large Language Models(LLMs) that aggregate and package knowledge, and when traditional curated web-sourced search results serve users better. 

An  experiment compared product searches using ChatGPT, Google search engine, or both helped us understand more about the compelling nature of generated responses. The experiment showed GenAI can speed up some explorations and decisions. We describe how search can deepen the testing of facts, logic, and context. We show where existing and emerging knowledge paradigms can help knowledge exploration in different ways.

 Experimenting with searches, our probes showed the value for curated  web  search provides for very specific, less popularly-known knowledge.  GenAI excelled at bringing together knowledge for broad, relatively well-known topics.  The value of curated and aggregated knowledge for different kinds of knowledge reflected in different user goals. We developed a taxonomy to distinguishing when users are best served by these two approaches.

\end{abstract}



\received{10 October 2024}

\maketitle

\section{Structure of Paper}
This paper proceeds along the following arc. The introduction presents a descriptive context for knowledge recording and exploration. 

We tested consumers in an experiment that involved a comparative car-buying exploration. This study included elements such as pre-test questionnaires, think-aloud protocols, and evaluations focusing on the time spent, the quantity and quality of alternatives considered, as well as post-test assessments.

A comparison of paradigms also inbcluded hundreds of probe queries using Google search and ChatGPT. These probes led to the creation and testing of 12 knowledge seeking personas, each with differing needs. This effort aimed to provide a representative exploration of various user types. For each persona, we proposed a range of experiences designed to highlight the differences between search engines and Generative AI (GenAI) tools. We include emblematic portions of a few of these instructive probes in this paper.

The quality of results from both approaches not only helped us delineate important differences between the two paradigms but also shed light on varying online knowledge needs. The paper concludes by proposing new user interface values for these emerging paradigms, tailored to the cognitive needs for information absorption.

\section{Introduction}
People spend their lives collecting knowledge to create their understanding of topics and areas that matter to them. They consider the building blocks of existing solutions and sometimes imagine new ones they would like to try.

Artificial Intelligence is a broad term that carries the implications of simulating human intelligence and has a history focused on computer reasoning, representation, and learning \cite{norvigbook}. The Machine Learning (ML) component has become a powerhouse for useful classification work across various industries. Databases and early search engines relied on humans to manually classify knowledge \cite{Shiftsinsearch,evaluatingsearch}. While today's search system classification still begins with human knowledge labeling, the most advanced ML technology layers deep structural connections to respond to complex queries \cite{efficientdeep}.

Search connects keywords using Boolean operators and presents lists of solutions relevant to knowledge and opportunities related to these queries \cite{googleSearchOperators}. "The search results draw from an AI representation of all \textit {curated} knowledge on the internet. Some search technologies utilize AI to generate web pages that serve as sources of information, as opposed to 'official' sources. For instance, search results for a restaurant often present a page compiled by the search engine, which aggregates information about the restaurant rather than relying solely on the restaurant's own description. Should we trust the restaurant to describe itself accurately, or could an \textit {aggregation} of various sources offer a more trustworthy account?"

A popular direction in AI involves training high dimensional predictive \textit{Transformer} to form language solutions . This approach leads to the creation of so-called Large Language Models(LLMs) \cite{StructureofATTENTionintranformer,Linguistics}. GenAI starts with human \textit{prompt} questions, including human-defined information and presentation goals. The LLM based solutions aggregate and \textit{package} it as a solution such as a story, poem, image, computer program, etc.\cite{unsettledbing}.   

In conversation, people are accustomed to not having all stories correct or complete. The value of discovering, evaluating, and expanding one's conceptions is best verified through traceable provenance. Whether in a street conversation or while viewing online information, source provenance must be made apparent to accumulate facts rather than rhetoric, thereby distinguishing reliable sources from fabricated stories. The search paradigm that has enabled us to find web information is valued for its ability to produce referenced responses.

\subsection{An Evolution of Knowledge Tools}
Technology has been aiding us in using language for a long time \cite{StructureofATTENTionintranformer}. While we still use graphite pencils to make marks, printers have allowed writers to create copies for distribution or editing. Displays now enable writers to view and modify writing at will.

Standardizing spelling and syntax made it easier for people to understand each other's writing. In 1604 Robert Cawdrey \cite{firstdictionary} standardized spelling in the first dictionary. Cut-and-paste and word completion features were introduced with Emacs, offering accuracy and flexibility in reorganizing thoughts.

Word-completion pull-down menus became a sought-after feature to speed up typing in Japanese word processors. Computer spell check became a valuable tool in the late 1970s. "Do What I Mean" (DWIM) added the ability to search for syntactic context and consistency in the 1980s \cite{Interlisp, teitleman}. 

In 1981, EPISTLE \cite{Text-Critiquing} was already using AI to improve authors' syntax \cite{MicrosoftWindows}.
Empathetic language responses were first demonstrated in the 1960s with Eliza \cite{weizenbaum1966eliza}. Systems like HearsayII in the 1970s allowed people to play chess by talking to a computer \cite{hearsay}.

The evaluation of ideas using computers has evolved. For a long time, creativity-enhancing software consisted of computer "outliners" and "mind maps." Now, we search the world's knowledge on Google.
Siri was a breakthrough in knowledge-based verbal question-answering [16, 17]. GenAI systems are combining user-focused text generation, as their responses prioritize solution flow and continuity over the correctness of the response answer [18]. The ability to guide GenAI with style and qualities of responses presents a completely new kind of interaction paradigm. The focus of requests and responses is on what should be presented and how. 

 \subsection{Knowledge Interaction Scenarios}
 Long before the introduction of GenAI, searching for results on the web was blurred. Search engines themselves reinterpret queries and produce results. Personal assistants like Siri also try to reduce questions from a person to an actionable query or action.
 We posed detailed questions to both the Alexa personal assistant and ChatGPT.

 When asked, "What is a quark?" both Alexa and ChatGPT give answers that seem to be condensed Wikipedia knowledge.
 Both of these platforms have limited corpora and knowledge bases compared to the web. Updating the indices of search systems is an up-to-the-second and ongoing comprehensive activity while updating GenAI systems is not yet as encompassing.  While Both have models of discourse, ChatGPT also has a model of persuasion, generating plausible results and filling in gaps with fanciful ideas. Alexa lacks the GenAI capability of filling in gaps with likely-sounding responses, as people do in conversation. So far, Alexa and ChatGPT both feel less reliable than internet searches for different reasons. Alexa finds an answer, while GenAI may try to prioritize and package top answers.

\subsection{New Scenarios for Knowledge Interaction}
Search has changed the way we find, use, and acquire things. The multiple goals of search have shifted its focus from finding a perfect answer to a negotiated set of results that both the search engine and the user interact with. Early purpose-built search engines were limited to finding web pages. Today's search engines serve multiple goals, providing marketing results, sales results, website results, video results, scholarly results, how-to results, and AI website-produced omposite results. A menu bar on Chrome allows Google Chrome users to focus on shopping, images, videos, news, maps, books, flights, and finance.
We have many goals in mind. The multifaceted goals of the search system mean that one searches and then finds themselves entangled in commercial opportunities that support the search business.
A more specialized search engine might help us focus. Perhaps the search engine shouldn't be promoting a new outfit, toolkit, or vacation opportunity when one is searching for a story's veracity and provenance. The CEO of a large search engine company was asked why Yahoo was presenting a married man with ads for breast enhancement and dating services. They said he said, "We tried to remove those ads, but the advertising opportunity generates a lot of revenue." Focusing these systems solely on users' goals will improve their reputation as well as productivity. With a subscription-based model, ChatGPT has so far avoided cluttering solutions with predatory results.

Starting in the 1990s, we were part of a class that projected a screen with a stream of Google search results for everything said. It was done in a yes-and-instigating manner that added to the conversation. Now we have much better tools. Let's explore some scenarios and the value they can bring.

\subsubsection{GenAI Can Feel Natural to Use} 
A few weeks after ChatGPT's initial release, we were amazed by  a group of octogenarians discussing their use of the technology. One of them recited a competent poem that ChatGPT had written for them in iambic pentameter, using Shakespearean language to discuss a current topic on their minds. 
\subsubsection{AI Can Provide Entertainment}
At a rain-forest resort in Sylhet, a group of people had asked ChatGPT to write a poem in the style of the most famous Bangladeshi poet. We watched in awe as of the groups  sang it as a song in Bengali.
\subsubsection{AI Can Generate Sophisticated Suggestions} 
We employed ChatGPT to author apologies for training examples about communication affect. Although the AI's suggestions became repetitive after a few sentences, they still helped people write training utterances almost 10 times faster.

A writer consulted ChatGPT for insights into disruptions following the Spanish flu pandemic, aiming to draw comparisons with expected disruptions after the COVID-19 pandemic. Upon fact-checking, the writer found that some of the AI-generated ideas were spurious. However, many were useful and even inspired the writer to come up with additional ideas. As a result, the time taken to complete the periodical article was significantly reduced.

\subsubsection{ Knowledge exploration needs guidance}
A friend asked ChatGPT about the best recording Shirley Temple made, and it said she never recorded. "It's worth noting that Shirley Temple, despite her fame as a child actress in the 1930s, retired from acting at the age of 22 and did not release any songs during her career." This statement, of course, was incorrect. The follow-up question aimed to provide more information, and ChatGPT then contradicted itself by noting her many real recordings.

\subsubsection{}{ GenAI Can Create}
We can search online support services like Stack Overflow to help us learn from other's programming examples. On the other hand, GenAI is now compiling all known code with tools like Copilot, assisting people in programming. Users can specify algorithms, the programming language to be used, and the desired output to receive working programs. By critiquing the result, they can prompt Copilot to fix bugs, change the approach, or even port the program to another language within seconds. Still, customizing these examples can be an extensive task that requires articulating specific requests as questions. However, programming complex tasks by asking questions can feel like trying to drive a car from the back seat.

While people are accustomed to conversation, search engines do not facilitate conversational or narrative interactions. When GenAI returns problematic responses and buggy programs, it initiates a conversation. One potential advantage of chat over search is that it encourages us to pause and evaluate the conversation. The clickbait nature of search tools may lead us to make hasty decisions without adequate consideration.

We can rely on chat systems for complete and organized responses, but people commonly state that (as with people) we can't trust them for truthfulness. Still, conversation may not be the most effective way to uncover facts or, to analyze and resolve bugs in a computer program. Critiquing and asking questions is different from creating. The critical inquiries of an art historian are not the creative acts of an artist. 

We spend our lives communicating to learn and accomplish tasks. One challenge is to develop a repertoire of knowledge and communication tools that are both valuable and productive. The search paradigm has been transformative, making older internet tools like Archie, Veronica, and WAIS forgettable. It provides us with access to the world's knowledge.

Search serves various knowledge needs. While newer versions now aggregate information about specific enterprises, such as restaurants or other businesses, they typically focus on displaying multiple alternative links that could be followed. Finding information is different from synthesizing a solution. The search paradigm is not designed to build knowledge, critically analyze discourse, or formulate solutions. The ease of accessing various forms of knowledge is now enhanced by the corpus of all digitized information. The AI-generated aspect added to search engines today might seem helpful, but represents a different kind of information not curated by people.  But how might we compare the curated knowledge access of search with the new GenAI paradigms?

Today's search engines are expected to respond appropriately to a wide range of informational needs. They aim to differentiate and present results tailored to a person's needs in well under a second. However, achieving this level of precision is challenging without understanding the context, background, and objectives of the request.

The GenAI paradigm draws from a portion of the same online knowledge that is available to legacy search systems. A key question arises: Can the automatically-generated narratives, which bring together parts of many disparate sources of knowledge, compete with the precise knowledge we all curate to be accessed by search engines? This question involves not just the quality and appropriateness of the knowledge but also its digestibility. While we often believe that we'll recognize what we want when we see it, the reality may be more complex.

The search systems of the past 20 years have enabled us to access a vast array of publicly presented information. However, the importance of structurally-coherent stories often outweighs the significance of knowing where and how we found something. Recent work has focused on demonstrating that helping people understand which GenAI results can be trusted may improve their decision-making \cite{spatharioti2023comparing, selker2023ai}. We designed an experiment  to see if the GenAI results might change the speed and way people make decisions.

A consumer purchasing experiment was designed to demonstrate how the different paradigms worked for people who believed in and also for people who didn't believe in GenAI.

\subsection{Consumer experiment method}
The primary aim of our experiment is to examine how using both legacy and new knowledge exploration platforms effects on complex, knowledge-based user decisions, such as buying a car. The study seeks to contrast how knowledge gathered from the web via legacy search engines and newly introduced GenAI aids consumers in this process. What are the most and least effective aspects of these online information access paradigms?

The study hypothesizes that a GenAI tool offers a spectrum of choices and hypothetical solutions, although these may not always be actionable. A combination of chat and search functionalities contribute to generating more actionable solutions, even though users may sometimes lack a systematic approach. When relying solely on search engines, users often find it challenging to initiate the process and may become bogged down in details, even though the results are more specific.

We began by recruiting individuals who are contemplating buying a car shortly and who reside in the U.S. The experiment was designed to last approximately one hour.

Before the experiment, participants answered a questionnaire to gauge their familiarity with search engines and GenAI chatbots. The questionnaire also aimed to explore their decision-making approach when it comes to buying a car.

Participants were required to complete three tasks.
Task 1: Use only a search engine for information gathering.
Task 2: Use only ChatGPT3.5 for the same purpose.
Task 3: Use both ChatGPT3.5 and a search engine, in an order of their choosing.
Each participant was tasked with researching a car purchase based on real-world scenarios and goals, such as finding the best car within a specific budget or comparing various car models while focusing on different lifestyle needs and preferences. They used Google search and ChatGPT3.5 tools to find several options, compare them, and narrow their choices down to two.

We employed screen recording, facial expressions, and verbal reactions to gain insights into their challenges, thought patterns, and decision-making processes.
The task was designed to be ecologically valid as a realistic, complex task. 

Recording the experience based on both how they acted and what they did helped us better understand the role and efficacy of online tools in complex decision-making scenarios like car purchasing.

\subsubsection{Response to the experiments}

We included three sections in our consumer experiment: pre-survey, three tasks, and post-survey.

In our pre-survey, we asked two questions to understand the participants' car purchase history and habits, All had purchased at least one car and 60 \%  said that they would buy a car based on the brand their family already owned, the other 40 \% rest would base it on what they found in a search.
 
Our purchasing experiment included three tasks. For each task, we asked the same question: "What might be different about using search or ChatGPT for finding cars?" [Figures 4, 7, and 10]. The reason we asked the same question was to compare the differences before, during, and after participants used ChatGPT3.5 for the search task.

Task One: Use only Google search to find a car.

Participants opened a browser and used the Google search box to learn about cars that meet the following requirements: a car suitable for an active lifestyle, that is safe, reliable, and also good for transporting kids.

They found several options[Figure 2], compared them, and identified differences. In the end, they narrowed their choices down to two[Figure 3].

\begin{figure}
  \centering
  \includegraphics[width = .75\linewidth]{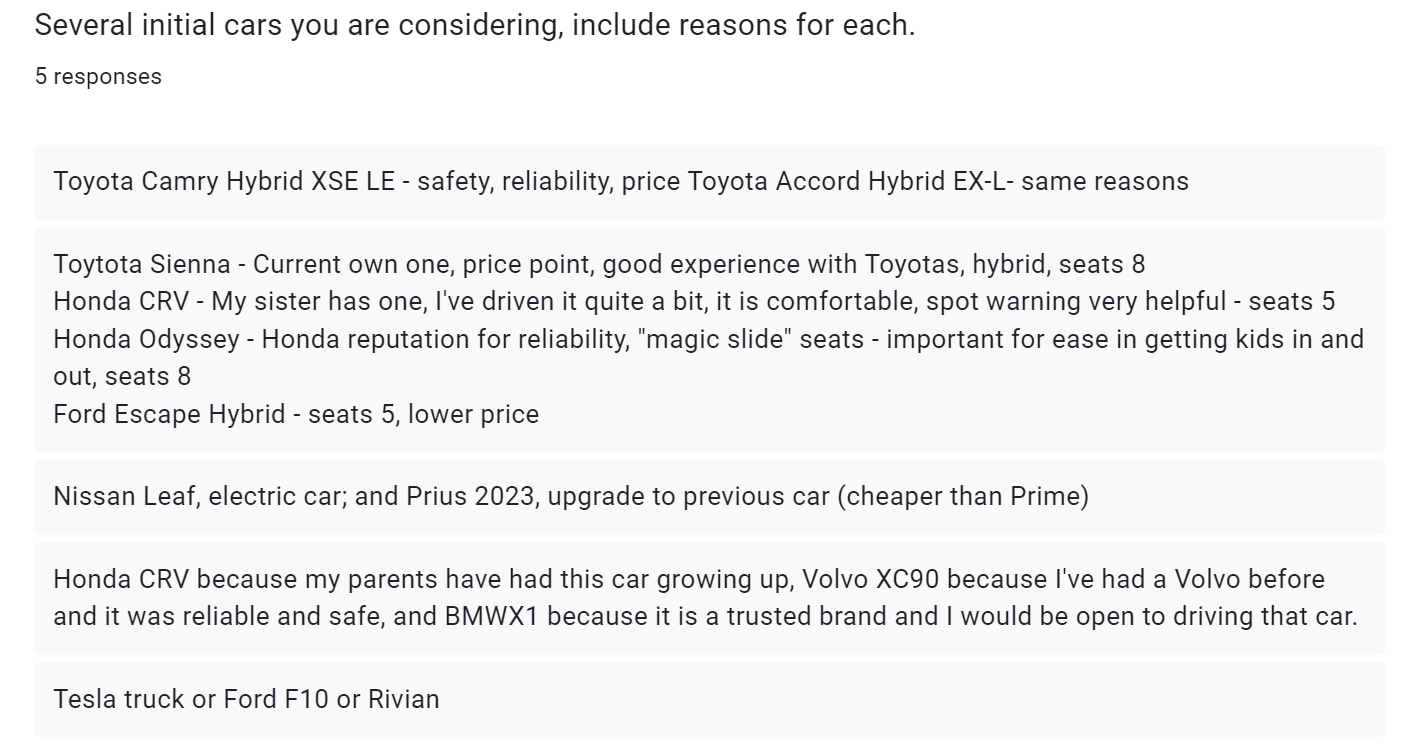}
  \caption{Google Task: Initial cars you are considering; include reasons for each.}
  \Description{Data: P1.Toyota Camry Hybrid XSE LE - safety, reliability, price 
Toyota Accord Hybrid EX-L- same reasons
P2. Toyota Sienna - Current own one, price point, good experience with Toyota's, hybrid, seats 8
Honda CRV - My sister has one, I've driven it quite a bit, it is comfortable, spot warning  is very helpful - seats 5
Honda Odyssey - Honda reputation for reliability, "magic slide" seats - important for ease in getting kids in and out, seats 8
Ford Escape Hybrid - seats 5, lower price
P3.
Nissan Leaf, electric car; and Prius 2023, upgrade to previous car (cheaper than Prime)
P4.
Honda CRV because my parents have had this car growing up, Volvo XC90 because I've had a Volvo before and it was reliable and safe, and BMW X1 because it is a trusted brand and I would be open to driving that car.
P5.
Tesla or Ford F150 or Rivian
  }
\end{figure}

\begin{figure}
  \centering
  \includegraphics[width = .75 \linewidth]{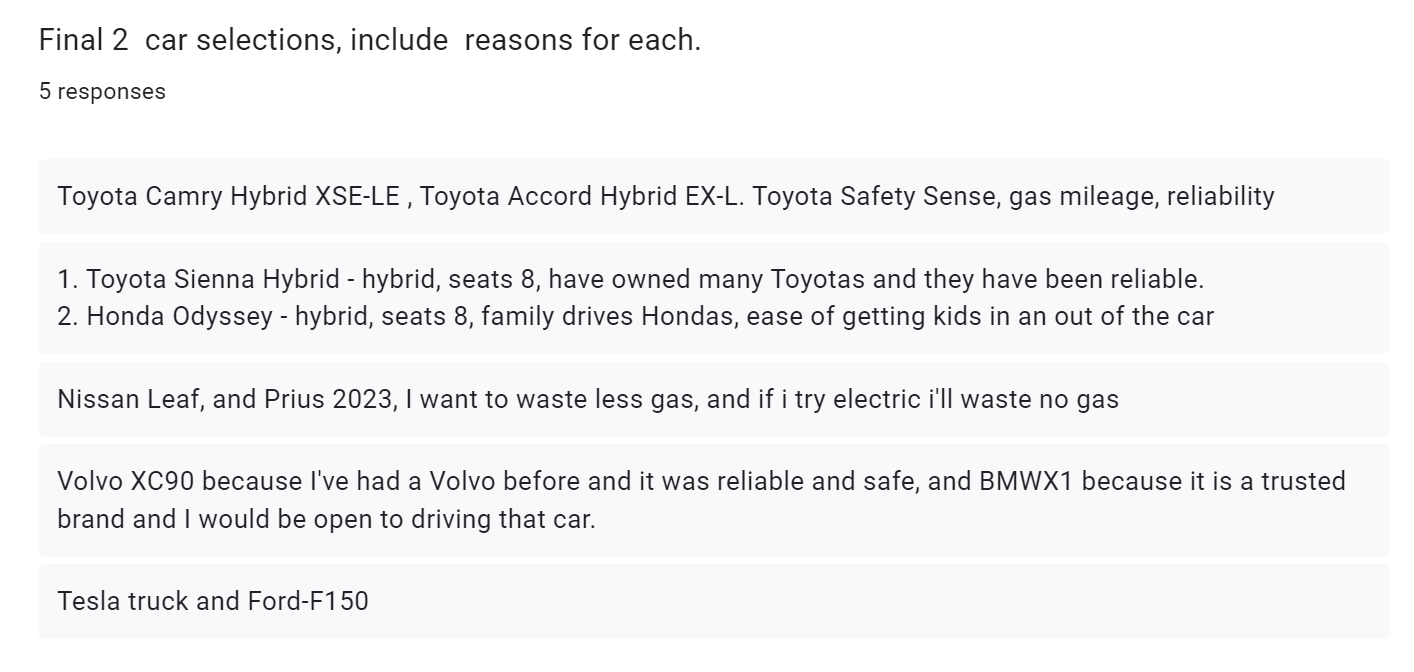}
\caption{Google Task: Final 2 car selections include reasons for each}
  \Description{Data: P1.Toyota Camry Hybrid XSE-LE , Toyota Accord Hybrid EX-L. Toyota Safety Sense, gas mileage, reliability,
  P2. 1. Toyota Sienna Hybrid - hybrid, seats 8, have owned many Toyotas and they have been reliable.
2. Honda Odyssey - hybrid, seats 8, family drives Hondas, ease of getting kids in an out of the car,
P3. Nissan Leaf, and Prius 2023, I want to waste less gas, and if i try electric i'll waste no gas,
P4. Volvo XC90 because I've had a Volvo before and it was reliable and safe, and BMWX1 because it is a trusted brand and I would be open to driving that car.
P5. Tesla truck and Ford-F150
  }
\end{figure}

 \begin{figure} 
  \centering
  \includegraphics[width = .75 \linewidth]{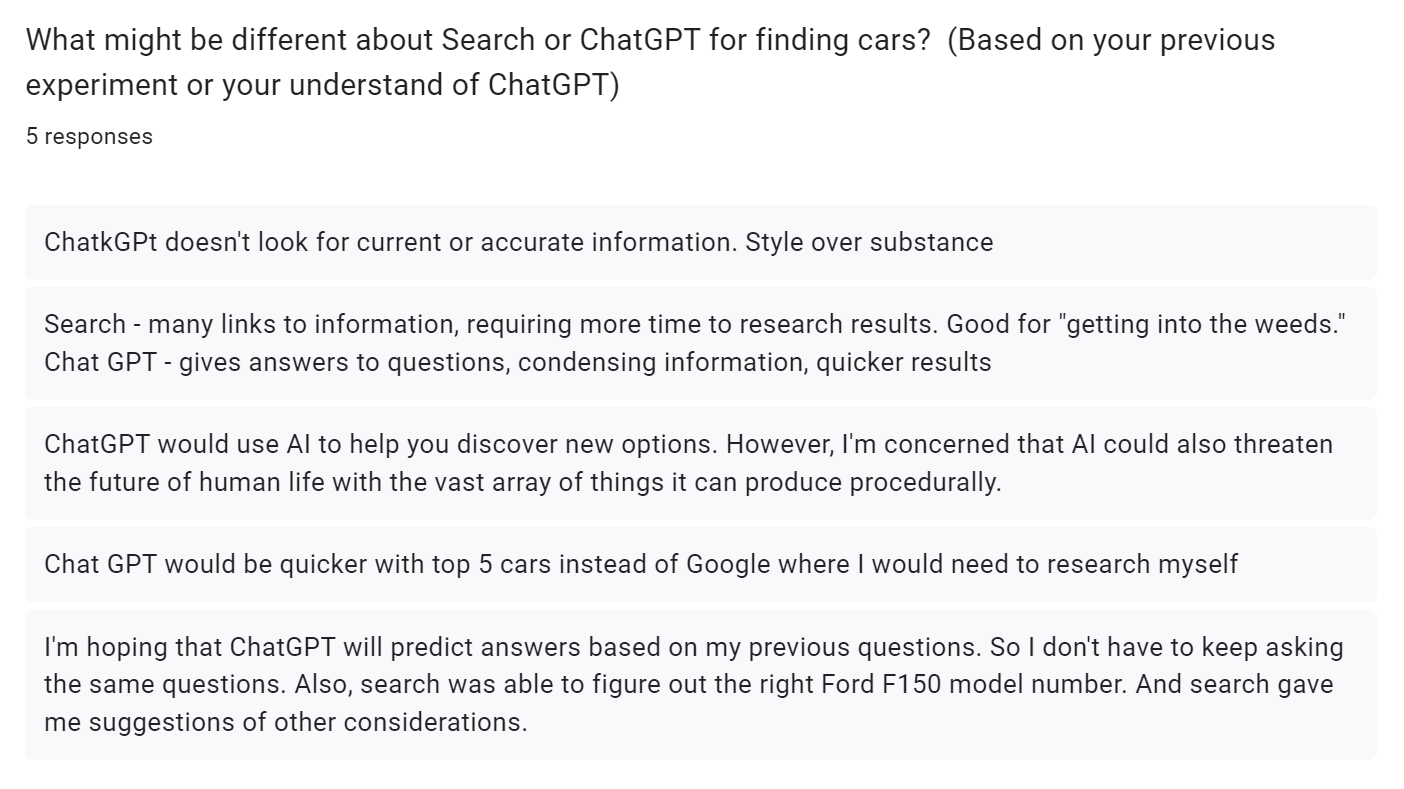}
\caption{Google Task: What might be different about Search or ChatGPT for finding cars?}
  \Description{Data: ChatGPT doesn't look for current or accurate information. Style over substance.
Search - many links to information, requiring more time to research results. Good for "getting into the weeds." Chat GPT - gives answers to questions, condensing information, quicker results
ChatGPT would use AI to help you discover new options. However, I'm concerned that AI could also threaten the future of human life with the vast array of things it can produce procedure.
ChatGPT would be quicker with the top 5 cars, instead of Google where I would need to research myself
I'm hoping that ChatGPT will predict answers based on my previous questions so I don't have to keep asking the same questions. Also, Search was able to figure out the right Ford F150 model number. And Search gave me suggestions of other considerations.
  }
\end{figure}

Task Two - Use only ChatGPT to find a car.

Participants used GenAI to find cars that meet the following requirements: suitable for an urban, luxury lifestyle, and good for transporting kids. They came across various alternatives [Figure 5] and assessed the distinctions between them. Finally, they settled on two final options [Figure 6].


\begin{figure} 
  \centering
  \includegraphics[width = .75\linewidth]{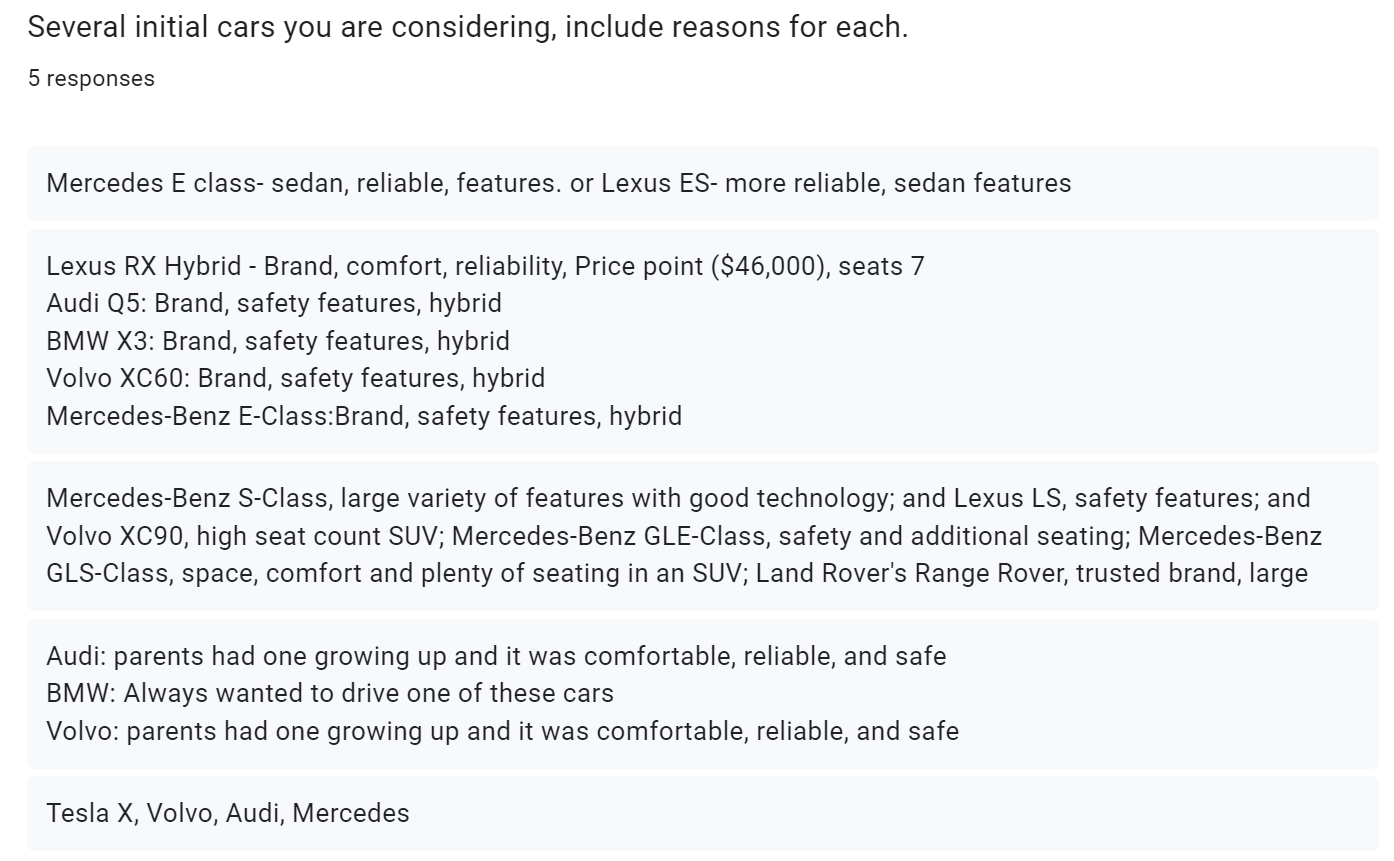}
  \caption{ChatGPT Task: Initial cars you are considering; include reasons for each.}
  \Description{Data: P1.Mercedes E class- sedan, reliable, features. or Lexus ES- more reliable, sedan features,
P2.Lexus RX Hybrid - Brand, comfort, reliability, Price point (\$ 46,000), seats 7
Audi Q5: Brand, safety features, hybrid
BMW X3: Brand, safety features, hybrid
Volvo XC60: Brand, safety features, hybrid
Mercedes-Benz E-Class:Brand, safety features, hybrid,
P3. Mercedes-Benz S-Class, large variety of features with good technology; and Lexus LS, safety features; and Volvo XC90, high seat count SUV; Mercedes-Benz GLE-Class, safety and additional seating; Mercedes-Benz GLS-Class, space, comfort and plenty of seating in an SUV; Land Rover's Range Rover, trusted brand, large,
P4. Audi: Parents had one growing up and it was comfortable, reliable, and safe
BMW: Always wanted to drive one of these cars
Volvo: Parents had one growing up and it was comfortable, reliable, and safe
P5. Tesla X, Volvo, Audi, Mercedes.
  }
\end{figure}

\begin{figure} 
  \centering
  \includegraphics[width = .75\linewidth]{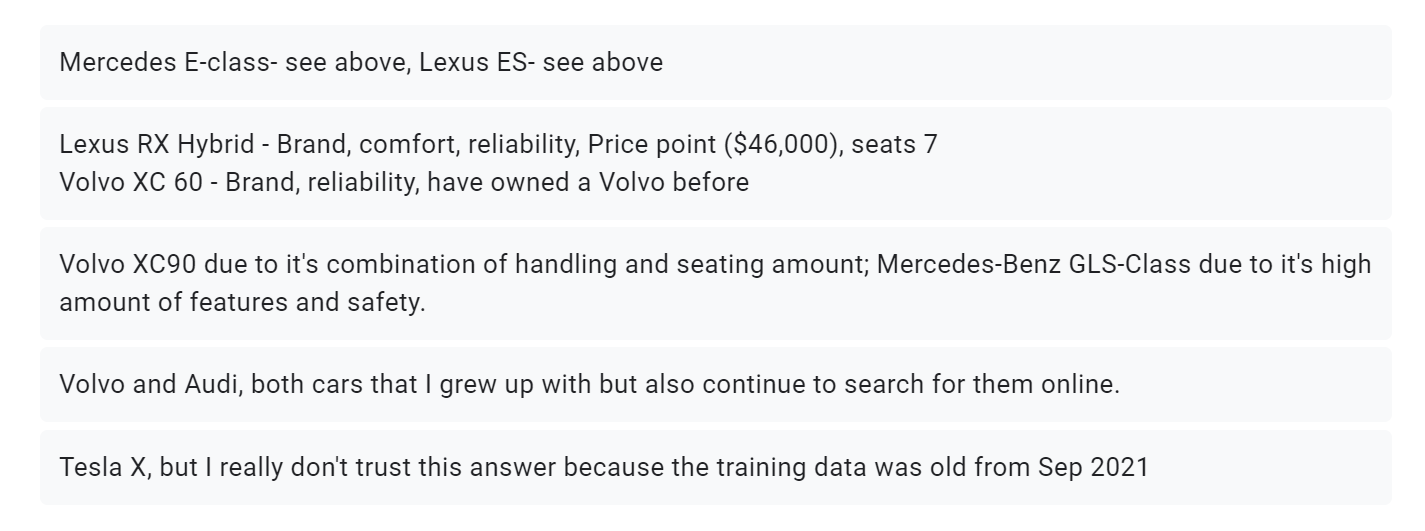}
\caption{ChatGPT Task: Final 2 car selections include reasons for each}
  \Description{Data: P1.Mercedes E-class- see above, Lexus ES- see above,
  P2. Lexus RX Hybrid - Brand, comfort, reliability. Price point ( \$ 46,000), seats 7.
Volvo XC 60 - Brand, reliability, have owned a Volvo before.
P3. Volvo XC90 - It's combination of handling and seating amount; Mercedes-Benz GLS-Class due to it's high amount of features and safety,
P4. Volvo and Audi - Both cars that I grew up with but also continue to search for them online,
P5. Tesla X - But I really don't trust this answer because the training data was old from Sep 2021.
}
\end{figure}

 \begin{figure} 
  \centering
  \includegraphics[width = .75\linewidth]{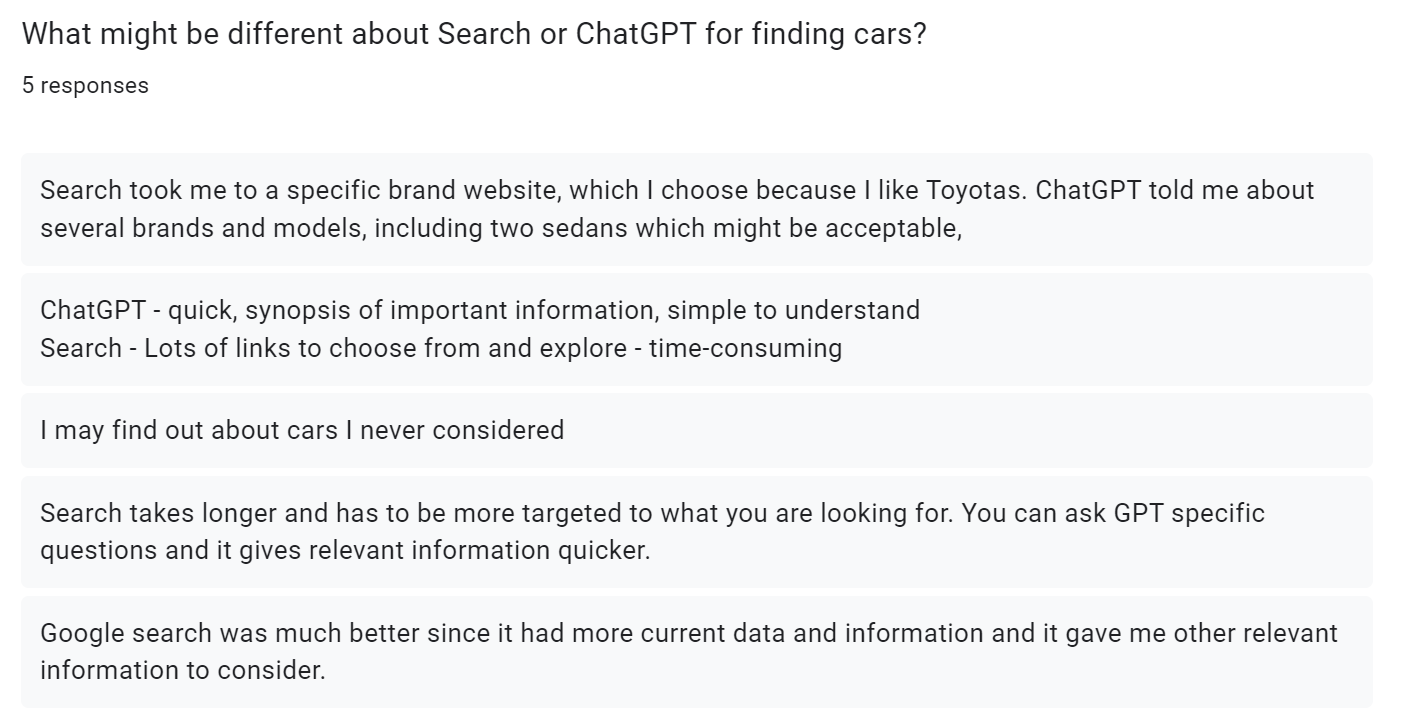}
\caption{ChatGPT Task: What might be different about Search or ChatGPT for finding cars?}
  \Description{ Data: P1.Search took me to a specific brand website, which I choose because I like Toyotas. ChatGPT told me about several brands and models, including two sedans which might be acceptable,
  P2. ChatGPT - Quick synopsis of important information, simple to understand. 
Search - Lots of links to choose from and explore - time-consuming
P3. I may find out about cars I never considered,
P4. Search takes longer and has to be more targeted to what you are looking for. You can ask GPT specific questions and it gives relevant information quicker,
P5. Google search was much better since it had more current data and information and it gave me other relevant information to consider.
  }
\end{figure}

Task Three - Use Both Google search and GenAI to find a car.

Participants checked details about cars that met the criteria for a rural lifestyle, fuel efficiency, low maintenance costs, and suitability for transporting kids—using both Google search and ChatGPT.
After discovering multiple options [Figure 8], they evaluated and distinguished the differences among them. Ultimately, they whittled their selections down to just two[Figure 9].

\begin{figure} 
  \centering
  \includegraphics[width = .75 \linewidth]{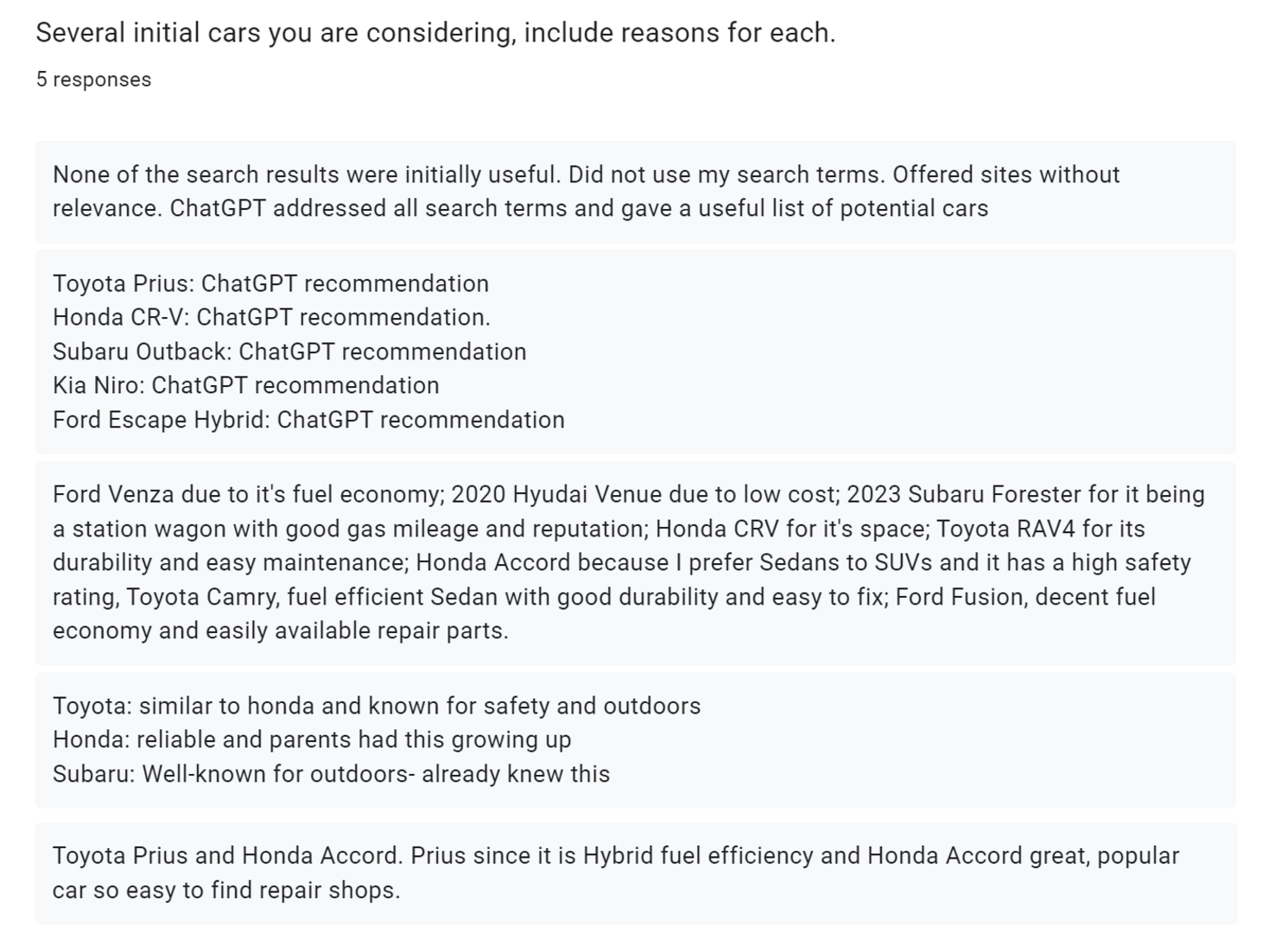}
  \caption{Google search and ChatGPT Task: Initial cars you are considering; include reasons for each.}
  \Description{Data: P1 None of the search results were initially useful. Did not use my search terms. Offered sites without relevance. ChatGPT addressed all search terms and gave a useful list of potential cars
P2.Toyota Prius: ChatGPT recommendation
Honda CR-V: ChatGPT recommendation,
P3. Subaru Outback: ChatGPT recommendation
Kia Niro: ChatGPT recommendation
Ford Escape Hybrid: ChatGPT recommendation
Ford Venza due to it's fuel economy; 2020 Hyudai Venue due to low cost; 2023 Subaru Forester for it being a station wagon with good gas mileage and reputation; Honda CRV for it's space; Toyota RAV4 for its durability and easy maintenance; Honda Accord because I prefer Sedans to SUVs and it has a high safety rating, Toyota Camry, fuel efficient Sedan with good durability and easy to fix; Ford Fusion, decent fuel economy and easily available repair parts,
P4. Toyota: similar to honda and known for safety and outdoors
Honda: reliable and parents had this growing up
Subaru: Well-known for outdoors- already knew this, 
P.5 Toyota Prius and Honda Accord. Prius since it is Hybrid fuel efficiency and Honda Accord great, popular car so easy to find repair shops.
}
\end{figure}

\begin{figure} 
  \centering
    \includegraphics[width = .75\linewidth]{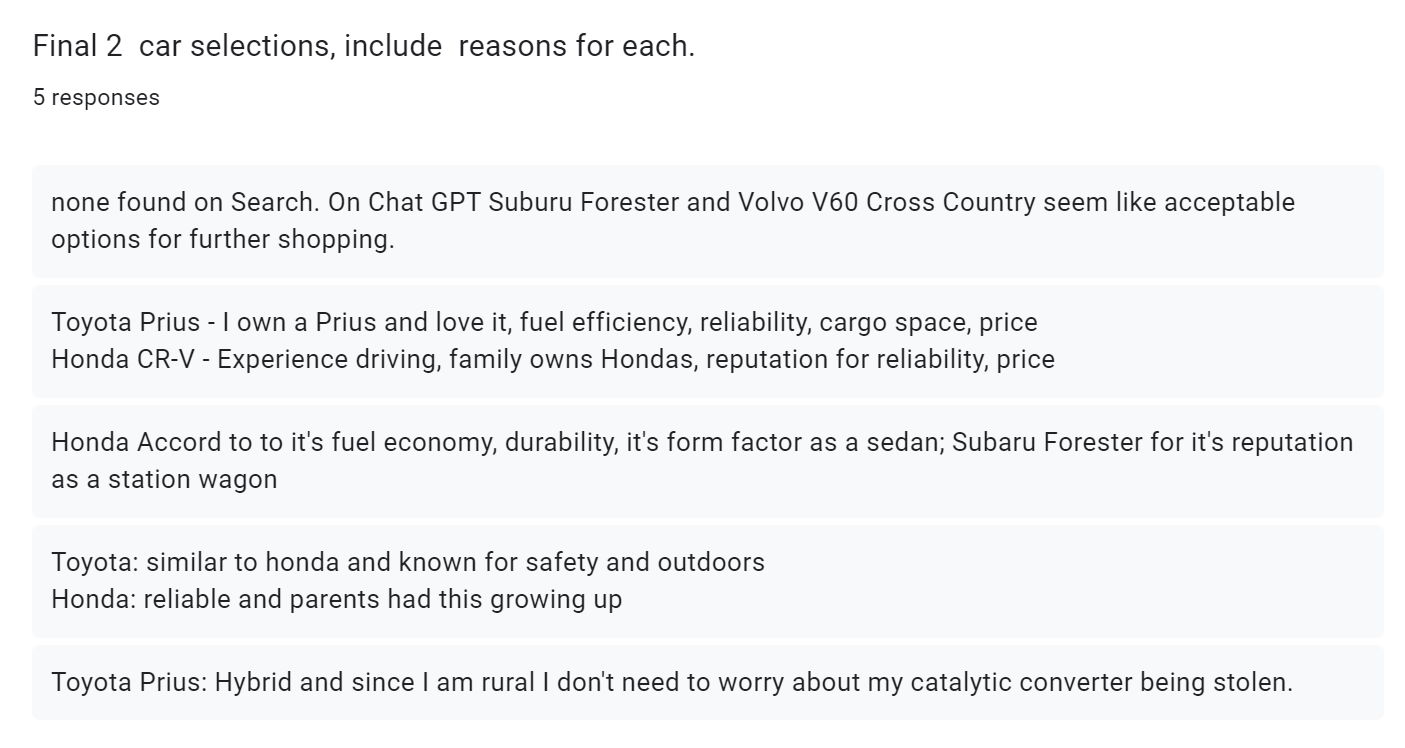}
\caption{Google search and ChatGPT Task: Final 2 car selections include reasons for each}
  \Description{Data:
P1. none found on Search. On ChatGPT Subaru Forester and Volvo V60 Cross Country seem like acceptable options for further shopping,
P2.
Toyota Prius - I own a Prius and love it, fuel efficiency, reliability, cargo space, price
Honda CR-V - Experience driving, family owns Hondas, reputation for reliability, price,
P3. Honda Accord: its fuel economy, durability, its form factor as a sedan; Subaru Forester for its reputation as a station wagon,
P4. Toyota: similar to Honda and known for safety and outdoors
Honda: reliable and parents had this growing up,
P5. Toyota Prius: Hybrid and since I am rural I don't need to worry about my catalytic converter being stolen.  
  }
\end{figure}

 \begin{figure} 
  \centering
  \includegraphics[width = .75\linewidth]{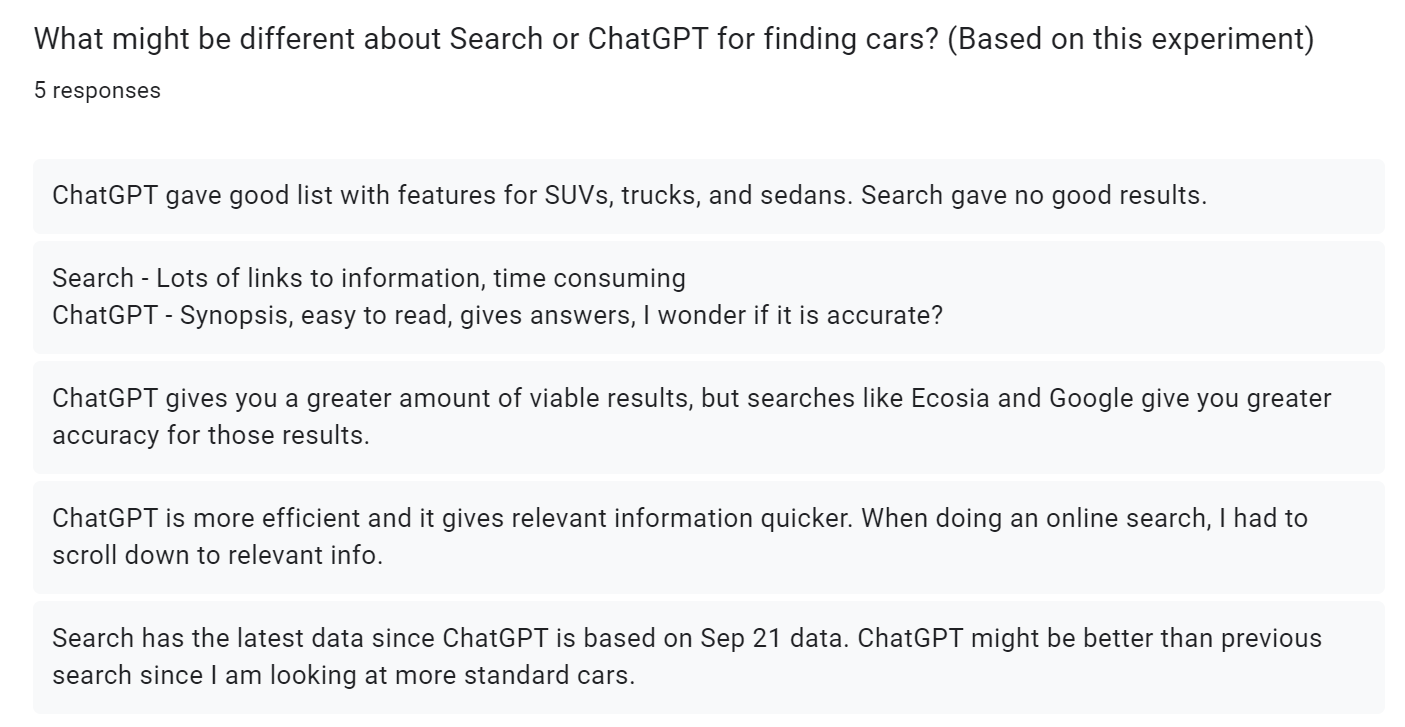}
\caption{Google search and ChatGPT Task: What might be different about Search or ChatGPT for finding cars?}
  \Description{Data: P1.ChatGPT gave good list with features for SUVs, trucks, and sedans. The search gave no good results,
  P2. Search - Lots of links to information,time-consuming
ChatGPT - Synopsis, easy to read, gives answers, I wonder if it is accurate?
P3. ChatGPT gives you a greater amount of viable results, but searches with Google give you greater accuracy for those results,
P4. ChatGPT is more efficient and it gives relevant information quicker. When doing an online search, I had to scroll down to relevant info,
P5. Search has the latest data since ChatGPT is based on Sep 21 data. ChatGPT might be better than previous search since I am looking at more standard cars.}
\end{figure}

Each participant's time spent on the task was documented [Figure 11].

 \begin{figure} 
  \centering
  \includegraphics[width = .75\linewidth ]{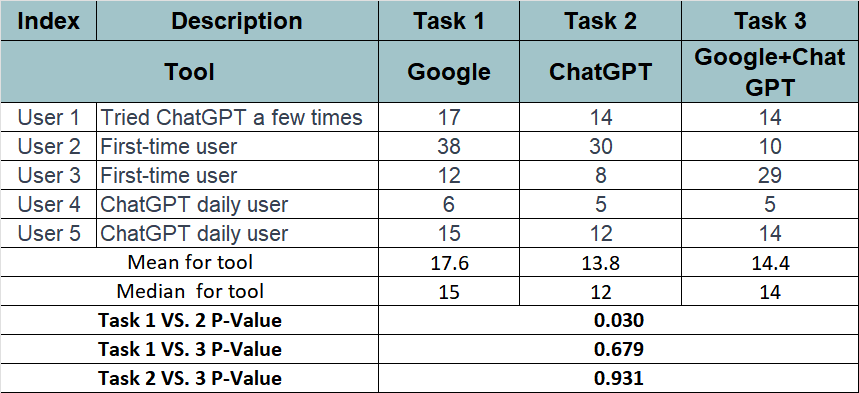}
\caption{The time spent on each user and task, p-value is calculated with the T-test}
  \Description{The table presents data from a study that measures how quickly different users completed three tasks using different tools: Google, ChatGPT, or a combination of Google and ChatGPT. 
  
Here's the breakdown:
Google
User 1 (Tried ChatGPT a few times): 17
User 2 (First-time user): 38
User 3 (First-time user): 12
User 4 (ChatGPT daily user): 6
User 5 (ChatGPT daily user): 15
Mean for Google: 17.6
Median for Google: 15
ChatGPT
User 1: 14
User 2: 30
User 3: 8
User 4: 5
User 5: 12
Mean for ChatGPT: 13.8
Median for ChatGPT: 12
Google+ChatGPT
User 1: 14
User 2: 10
User 3: 29
User 4: 5
User 5: 14
Mean for Google+ChatGPT: 14.4
Median for Google+ChatGPT: 14
p-Values (Statistical significance)
Task 1 vs. Task 2: 0.030
Task 1 vs. Task 3: 0.679
Task 2 vs. Task 3: 0.931}
\end{figure}

Following the completion of the three tasks, we solicited participants' opinions on the tools through a post-survey. What drawbacks do you perceive in ChatGPT when compared to Google search[Figure 12]? What benefits do you see in using ChatGPT over Google search[Figure 13]? How would you rate the ease or difficulty of locating the information you required using both Google search and ChatGPT[Figures 14]? Another question delves into whether ChatGPT was helpful[Figure 15], and the last question explored the frequency with which they utilize ChatGPT[Figure 16].

\begin{figure} 
  \centering
  \includegraphics[width = .75\linewidth]{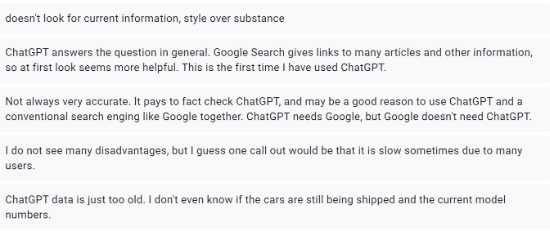}
\caption{What do you think are the disadvantages of ChatGPT compared to Google search?
}
  \Description{Data: P1. doesn't look for current information, style over substance,
  P2.
ChatGPT answers the question in general. Google search gives links to many articles and other information, so at first look seems more helpful. This is the first time I have used ChatGPT,
P3.
Not always very accurate. It pays to fact check ChatGPT, and may be a good reason to use ChatGPT and a conventional search engine like Google together. ChatGPT needs Google to check accuracy, but Google doesn't need ChatGPT for such checks,
P4.
I do not see many disadvantages \textit{(to ChatGPT)}, but I guess one call out would be that it is slow sometimes due to many users,
P5.
ChatGPT data is just too old. I don't even know if the cars are still being shipped and the current model numbers.
  }
\end{figure}

\begin{figure} 
  \centering
  \includegraphics[width = .75\linewidth]{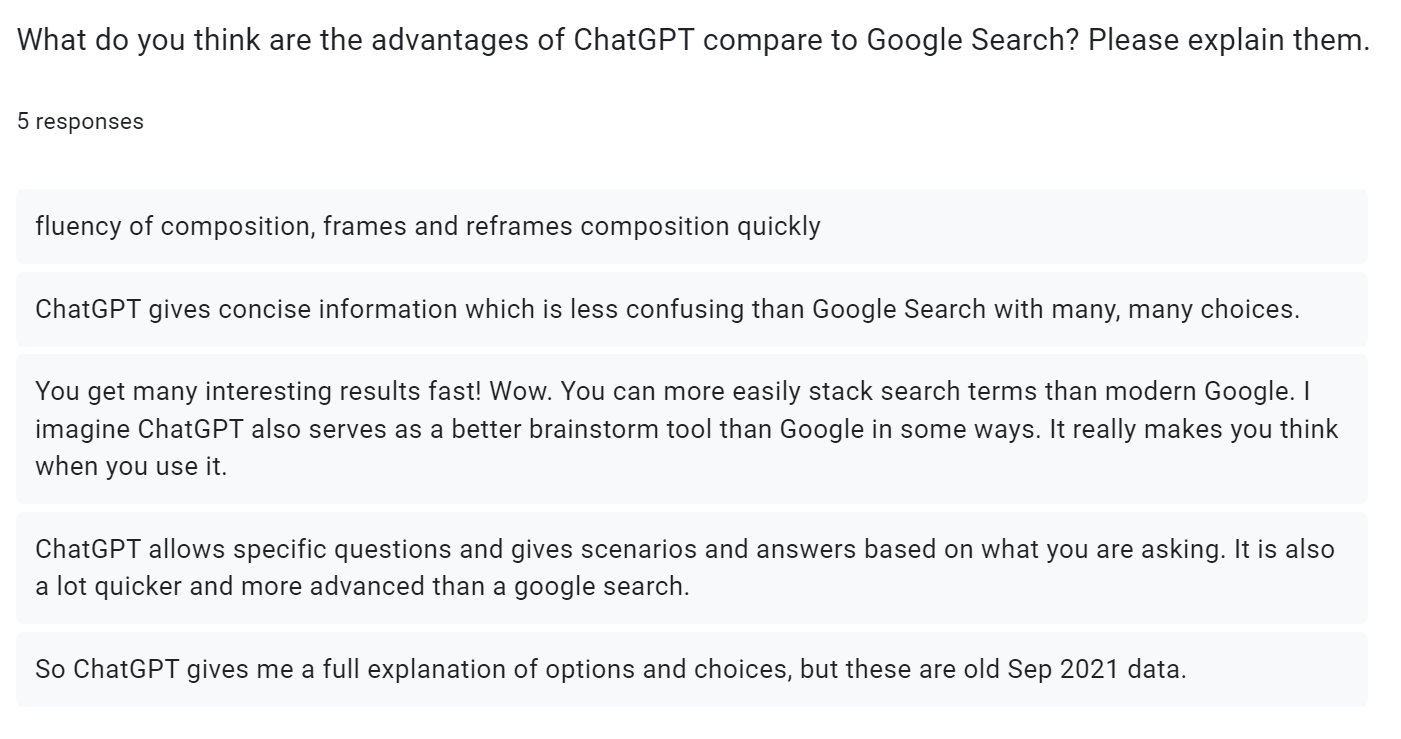} 
\caption{What do you think are the advantages of ChatGPT compared to Google search? }
  \Description{Data: P1. fluency of composition, frames and reframes composition quickly,
P2.ChatGPT gives concise information which is less confusing than Google search with many, many choices.
P3. You get many interesting results fast! Wow. You can more easily stack search terms than modern Google. I imagine ChatGPT also serves as a better brainstorm tool than Google in some ways. It really makes you think when you use it.
P4.ChatGPT allows specific questions and gives scenarios and answers based on what you are asking. It is also a lot quicker and more advanced than a Google search.
P5. So ChatGPT gives me a full explanation of options and choices, but these are old Sep 2021 data.
}
\end{figure}


 \begin{figure}
   \centering
\includegraphics[width= .75\linewidth]{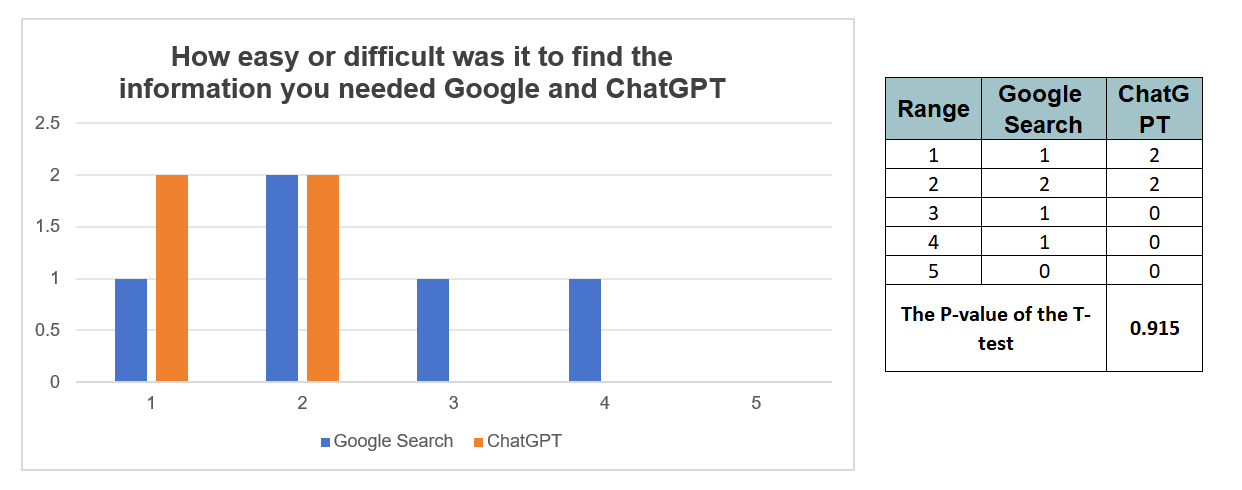}
  \caption{How easy or difficult was it to find the information you needed Google and ChatGPT}
  \Description{The content represents a table comparing the ease or difficulty with which participants found information on Google search and ChatGPT. In this survey, 1 is marked as the easiest to find information and 5 is the most difficult.

For Google search:

1 participant found it to be very easy (score of 1).
2 participants found it somewhat easy (score of 2).
1 participant found it of average ease (score of 3).
1 participant found it somewhat difficult (score of 4).
No participants found it to be very difficult (score of 5).
For ChatGPT:

2 participants found it to be very easy (score of 1).
2 participants found it somewhat easy (score of 2).
No participants found it of average ease (score of 3).
No participants found it somewhat difficult (score of 4).
No participants found it to be very difficult (score of 5).}
\end{figure}


 \begin{figure}
  \centering
\includegraphics[width = .75\linewidth]{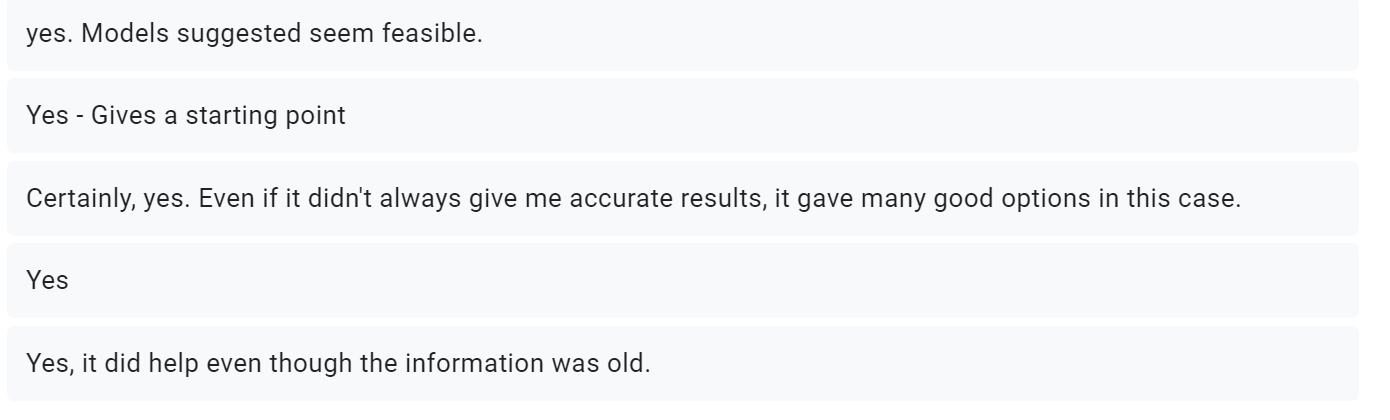}
\caption{Did ChatGPT help?}
  \Description{ Data: P1.yes. The models suggested seem feasible.
Yes - Gives a starting point
Certainly, 
P2.yes. Even if it didn't always give me accurate results, it gave many good options in this case.
P3. Yes,
P4.Yes, it did help even though the information was old
}
\end{figure}

 \begin{figure}
  \centering
\includegraphics[width= .75\linewidth]{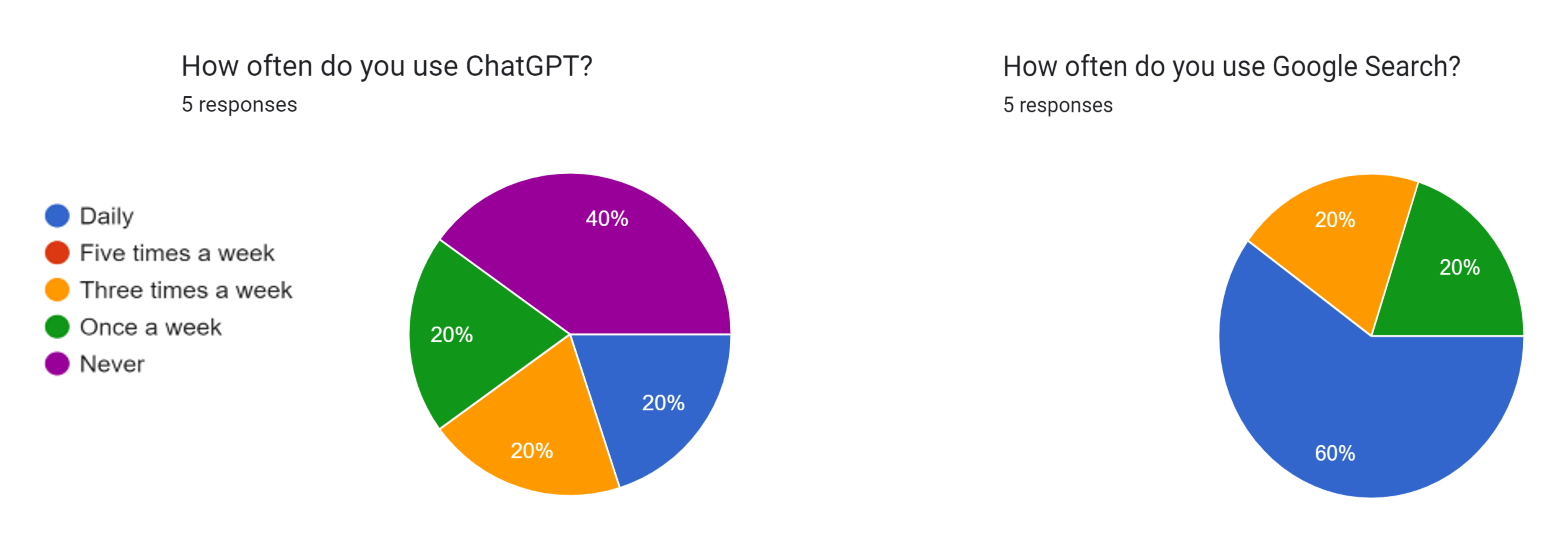}
\caption {How often do you use ChatGPT and Google search?}
  \Description{How often do you use ChatGPT and Google search?}
\end{figure}

\subsubsection{Interpretation of consumer experiment }
\paragraph{Pre-survey findings}
All participants had prior experience purchasing one or more vehicles.
Participants were very focused on their preconceptions about different car brands and types. Even when viewing new information, they relied on their own research and familial car ownership when proposing a car purchase .

Through the think-aloud process, we recorded that while all participants were aware of ChatGPT, two were first-time users who had very negative expectations about what the experience would be like.

\paragraph{Participant feedback on three tasks and time spent[Figures 4,7, and 10]:}
ChatGPT: Focuses on stylistic presentation rather than on current or accurate information.
Search vs. ChatGPT: Google search requires several passes for more in-depth research, whereas ChatGPT provides concise answers upon the first request.

AI Concerns: While ChatGPT can suggest new options, concerns about the long-term societal impact of AI were raised.

Efficiency: ChatGPT can quickly list top options, saving user organizing effort and time.

Personalization: ChatGPT offers personalized responses, while search engines can provide specific models and additional suggestions.

Range of Options: ChatGPT provided a more comprehensive list in one answer.

Time and Effort: It took several steps to obtain the needed information using a search engine; ChatGPT offered summarized information in a single step.

Accuracy Concerns: Questions were raised regarding ChatGPT's accuracy.

Quality vs. Quantity: ChatGPT offered more options, but search engines provided more accurate results.
Data Currency: Search engines offered more up-to-date information.
Observations on the use of time on each task:
Experienced Users were Faster: Users who were familiar with ChatGPT (Users 1, 4, and 5) generally completed tasks faster than those who were not (Users 2 and 3) [Figure 11].

Daily users are most efficient: Users 4 and 5, who are daily users and 4.0 subscribers, had the fastest times across all tasks, indicating that familiarity and subscription level may correlate with efficiency [Figure 11].

Summary of Statistical Analysis on Time Taken for Tasks 1, 2, and 3
Task 1 and Task 2 Comparison:
Null Hypothesis: A t-test did not show a significant difference in time taken between using Google search (Task 1) and ChatGPT (Task 2) for the tasks.

Result: A p-value of 0.030 was obtained,  This allows us to reject the null hypothesis, suggesting that there is a statistically significant difference in the time taken to complete tasks using Google search and ChatGPT.

Task 1 and Task 3 Comparison:
Null Hypothesis: There is no meaningful difference in the time required to complete Task 3 (utilizing both Google search and ChatGPT) as compared to Task 1 (utilizing only Google search).

Result: A p-value of 0.68 was recorded, which significantly above significance threshold. This outcome leads us to not reject the null hypothesis, our data doesn't provide enough evidence to indicate a difference in the time required for Task 3 versus Task 1.

Task 2 and Task 3 Comparison:
Null Hypothesis: There is no significant difference in the time taken to complete Task 2 (ChatGPT) and Task 3 (Google search and ChatGPT).

Result: A p-value of 0.93 was obtained,  This shows insufficient evidence to indicate a significant difference in time taken between Task 2 and Task 3.

Overall Conclusion:
There is a statistically significant difference in the time taken to complete tasks when comparing Google search and ChatGPT (Task 1 vs. Task 2).

However, our data does not show evidence for a statistical time difference for  Google search and ChatGPT (Task 3), in time taken compared to using either Google search (Task 1) or ChatGPT (Task 2) alone[Firgure 11].

\paragraph{Post-Study Survey:}
Advantages of ChatGPT: All participants found ChatGPT to be fluent, concise, and efficient, offering a tailored search experience and aiding in brainstorming and critical thinking.

Disadvantages of ChatGPT: While ChatGPT was useful for quick queries, it lacked depth, accuracy, and current information compared to Google search.

Ease of use in ChatGPT: On a scale of 1 to 5, with 1 being the easiest and 5 being the hardest, participants rated the ease of finding information on ChatGPT between 1 and 2.

Ease of use in search: On the same scale, Search scored between 1 and 4.

Null Hypothesis: There is no significant difference in the ease or difficulty ratings between Google search and ChatGPT.

A p-value of 0.915 is quite high, well above the commonly used significance level of 0.05. It fails to reject the null hypothesis.

In our data, it appeared that ChatGPT is easier to use than Google search. However, we would need at least need more subjects to  see a significant difference in the p-value. 

All but one participant expressed a preference for ChatGPT following the completion of the three tasks.
One participant found search accuracy much more important than its presentation.
Based on the data and feedback from our pilot study, it seems that the experiment largely supports our initial hypothesis. Here's a breakdown:
The docent administered the experiment with a uniform script for each user. As well as the results of the pretest, searches, and post-tests, recorded observations were made of each participant. As the data above shows, even the participants that were not oriented towards using GenAI, were able to appreciate it, productively use it, and integrate its results. The typical time to go through the three tasks was one hour. One participant who was very used to knowledge exploration did everything in 20 minutes. Although everyone showed their prior biases about car brands in their reactions to GenAI suggestions, one emblematic participant was incredulous at results that didn't match their brand loyalty and prior beliefs. Still, that participant rated GenAI as valuable to their exploration.

\subsubsection{Consumer Experiment  Observations}
GenAI packages and organizes a spectrum of choices and hypothetical solutions, but these may not always be actionable.

Supported: Participants found that ChatGPT offered a broad range of options but raised concerns about its accuracy. This suggests that while ChatGPT can generate a variety of choices, these may not always inspire action.

A combination of chat and search functionalities helps in generating more actionable solutions.

Supported: Participants who used both Google search and ChatGPT found that they could obtain a more comprehensive list of options. This suggests that the combination does indeed provide more actionable solutions.

Users may sometimes lack a systematic approach.

Partially Supported: While the study doesn't directly address this, the feedback that ChatGPT helps in "brainstorming and critical thinking" suggests that users find the chat format helpful in structuring their approach.

When relying solely on search engines, users often find it challenging to initiate the process and may get bogged down in details.

Partially Supported: Participants noted that Google search requires more in-depth research, which could be interpreted as getting "bogged down in details." When  then does search give us what we want and when does it bog us down in the details?

We were led to consider and try to  classify various types and uses of knowledge that influence how we can best access information through both search and GenAI.  We then observe how people react to various parts of the same question when posed within search-based versus generative paradigms.


\subsection{Evaluating search and GenAI results for 12 personas}
Our first set of evaluations focused on exploring various approaches to knowledge exploration, leading us to create 12 distinct search information exploration personas. For each persona, we began with assumptions about the utility of both curated and aggregated knowledge. These assumptions were then tested through specific search scenarios to evaluate the effectiveness of different types of responses. Due to space limitations, we are unable to present all the scenario results or delve into the setup, data, and interpretation in detail within this paper. These additional insights are included in Appendices A and B. 

In the following sections, we present the personas, hypotheses, the two most useful prompt topics, and our interpretations.
Format of data  format below:

\begin{enumerate}
\item \textbf{Persona name} hypothesis text
\begin {itemize} 
\item Topics: useful prompt topics.
\item Interpretation of if \textbf{search} or \textbf{GenAI} gave better results. 
\end {itemize}
\end {enumerate}

\begin{enumerate}

\item \textbf{A Do It Yourself person} describes an individual when they are  attempting to fix or build things on their own, without professional expertise or tools. They often search for how-to guides or tutorials to solve problems without needing to consult an expert or enroll in a course. Such behavior presents monetary opportunities in the realms of education, materials, and services.
\begin {itemize} \item
Topics 1. garden shed, 2. car not starting
\item \textbf{Search} included steps, images, and videos. 
\end {itemize}
\end {enumerate}

\paragraph{Persona prompts best served by \textbf{Search}}
\begin{enumerate}
\item \textbf{A recreational intellectual} describes an individuals who seeks to stay updated in a particular field, even if they are unlikely to contribute original work to it. Their searches typically focus on well-known facts, whether trivial or more substantive. Since these intellectual pursuits are often undertaken for entertainment, search engines could also appropriately suggest related entertainment options, such as food and drink.

\begin {itemize} \item
Topics 1. trivia, 2. cultures
\item \textbf{Search} tended to provide more complete knowledge. 
\end {itemize}
\item \textbf{A Do It Yourself person} describes an individual when they are  attempting to fix or build things on their own, without professional expertise or tools. They often search for how-to guides or tutorials to solve problems without needing to consult an expert or enroll in a course. Such behavior presents monetary opportunities in the realms of education, materials, and services.
\begin {itemize} \item
Topics 1. garden shed, 2. car not starting
\item \textbf{Search} included steps, images, and videos. 
\end {itemize}
\item \textbf{A hobbyist} applies to people who are actively participating in a specific nonprofessional interest, often in the company of others who share the same passion. These individuals frequently seek out communities and avenues to enhance and gain visibility for their interests. Many of their informational needs, especially those related to visual and video content, can be easily met through standard search engines.
\begin {itemize} \item
Topics 1. find stamp collectors 2. bird watching places
\item \textbf{Search} shared more details about collectors and ways to go bird watching. 
\end {itemize}
\item \textbf{A constructor} refers to individuals focused on creating something new or impressive, as cited in \cite{Cyberiad}. These people seek methods, materials, and analyses to support their creative endeavors. Standard search tools prove effective in meeting these needs.

\begin {itemize} \item
Topics 1. energy efficient construction techniques, 2. best coding practices 
\item \textbf{Search} gave more diverse results.
\end {itemize}
\item \textbf{A domain expert} is an Individual who has extensively researched a specific topic. While similar to scientists in their depth of knowledge, these individuals are not necessarily interested in conducting experiments. Instead, they seek comprehensive and specialized information in their area of expertise. 
\begin {itemize} \item
Topics 1. Advanced chess strategies, 2. advanced organic gardening techniques
\item \textbf{Search} provided useful videos.
\end {itemize}
\end{enumerate}

\paragraph{Persona prompts served well with both \textbf{genAI} and \textbf{Search}}

\begin {enumerate}
\setcounter{enumi}{5}

\item \textbf{The professional} is an individual who uses validated practices and tools in their jobs. Their searches are often contextual and tailored to specific needs and situations.
\begin {itemize} \item
Topics 1. current accounting jobs, 2. marketing courses or certificates 
\item The two aradigms gave similar results, but \textbf{GenAI} prioritized results without advertising prejudice. 
\end {itemize}
\item \textbf{The merchant} describes an individual engaged in buying or selling goods and services. These individuals aim to discover marketplaces and potential customers interested in their offerings. Specialized search websites are often effective in quickly providing the information and interactions needed to achieve these goals. 
\begin {itemize} \item
Topics 1. fashions that will sell, 2. market for handcrafted furniture
\item \textbf{Search} gave quick access to examples, \textbf{GenAI} categorized well.
\end {itemize}
\item \textbf{The consumer} describes an individual focusing on acquiring goods or services. They assess both known and unknown needs, exploring opportunities to both envision and fulfill them. While standard search sites often rely on basic collaborative filtering algorithms, specialized websites can be more efficiently accessed through targeted searches.
\begin {itemize} \item
Topics 1. exploring a getaway trip, 2. best smartphone
\item \textbf{Search} was better for viewing specific places and products. \textbf{GenAI} organized choices from an article about phones.
\end {itemize}
\end{enumerate}
\paragraph{Persona prompts best served with \textbf{genAI} responses }
\begin {enumerate} 
\setcounter{enumi}{8}
\item {\textbf{A fact checker} describes an individual when they are dedicated to verifying the source and accuracy of information. Unlike consumers or merchants, they are not focused on purchasing items. In this context, provenance is paramount; the information must be both accurate and traceable.}
\begin {itemize} \item
Topics 1. reporter, 2. historian 
\item \textbf{GenAI} was more direct and focused. 
\end {itemize}
\item \textbf{A scientist} describes individuals when they are engaged in the pursuit of knowledge to formulate and test hypotheses within a specific field. Their primary objective is not limited to confirming existing information but extends to seeking related knowledge from established experts within their domain, as well as from emerging researchers and unconventional thinkers. Colleague interactions and thorough examination of scientists' published works and bibliographies serve as their customary sources of information.
\begin {itemize} \item
Topics 1. Geneticist, 2.Astrophysics 
\item \textbf{GenAI} made it easier to browse more references. 
\end {itemize}
\item \textbf{A student} refers to individuals when they are in the process of acquiring knowledge and skills, aiming to gain an understanding of various subjects and techniques. Students often  demonstrate their proficiency through test taking and demonstrating knowledge in projects. GenAI has proven to be immensely valuable in assisting students in achieving their educational objectives by providing support and solutions in this context

\begin {itemize} \item
Topics 1. explain sorting algorithms, 2. artistic styles of the renaissance.  
\item \textbf{GenAI} gave comprehensive results that helped and followup questions that were instructive.
\end {itemize}
\item \textbf{A communicator} describes individuals  who are engaging in the process of scripting presentations. They assemble content with the intent to share, educate, or document information. GenAI is revolutionizing this domain with its innovative paradigm for creating such materials.

\begin {itemize} \item
Topics 1. blog writing support, 2. writing a friendly manual for software
\item \textbf{GenAI} Chat supported user writing and style goals!
\end {itemize}
\end {enumerate}

\subsubsection{Persona experiment discussion}

The topics revealed several differences for what kinds of prompts were best served by \textbf{search} curated web knowledge and which were better served by packaged \textbf{GenAI} aggregated knowledge. An interesting finding is that while there are many knowledge needs that are better served with \textbf{search} there are huge benefits to \textbf{GenAI} for communicating ideas to people.

\begin {itemize}
\item For very specific requests such as how to replace a fender-well in a 2010 Prius
\textbf{GenAI} gave a generic list of steps and tools that weren't appropriate; \textbf{search} gave specific results with video demonstrations for our specific car. 

Note: We replaced a fender well in our driveway without a jack in under 30 minutes using a screwdriver, and a wrench the responses asked for but knew we needed things both online information systems had missed: several new clips and a knife to pop the tough old ones out. 

\item Explorations for commonly-available information yielded similar results with the two approaches but GenAI will package it in a comprehensive form that is easy to scan.

\item As Maps is closely integrated with search systems, one might expect search to perform better with spatial responses. This was noted by the consumer above who was looking for a getaway. 
Still, the GenAI's ability to assemble responses from various sources gave excellent answers.

\item \textbf{GenAI} results were easier to look through as they presented in narrative form and didn’t have the clutter of advertising and other search goals not intended in the question. 

\item \textbf{GenAI} had the added advantage of showing a prioritization of results explicitly.

\item \textbf{GenAI} had the ability to continue to follow, expand and focus a question is demonstrated well in examples such as a student using it as a tutor.
\item \textbf{GenAI} results coming as a presentation is demonstrated well for a communicator whose result will be a presentation.
\end{itemize}
Generalized search engines like Google have been trying to cover all the uses a person might have for it, reducing the need for specialized tools. Trying to cater to any information need, they produce lists and web pages. Still, special-purpose search engines or search engine settings can be useful for looking through refereed articles, different media, visualization, domain access, and specialized information. 

The emerging \textbf{GenAI} paradigm lets a user add any amount of context for a query naturally as they define and redefine their prompt request. For many people, the discursive input of ChatGPT is turning out to be easier to master than the Boolean-esque languages search engines depend on for user input. Users typically formulate a search for a question with keywords. GenAI tools encourage people to formulate questions as a narrative, including information and presentation constraints—an act of exploration that might make us more open to diverse answers.
 
\subsubsection{Why GenAI Appeals to Users}
From calculators, interfaces like spreadsheets, and modeling systems, to query languages, digital assistants, bots, and search, computers answer our questions in various ways. Now, we are enamored with the ease of solving problems using GenAI. What are the key elements that make GenAI more accessible and appealing compared to other question-answering paradigms?
\begin{itemize}
\item Is it because proposed solutions are easier to follow than a taxonomic list?
\item Do people take naturally to the output because it shows the confidence of a proponent presenting a solution? 
\item Is it because it answers with proposed solutions, not facts?
\item Is it the way it adds in connective plausible ideas that give smooth transitions? Are these “hallucinated” fantasies plausible ideas useful for reinforcing known facts? In making things fit together, people also add such plausible but inaccurate ideas to connect facts to and get away with them in promoting a narrative.
\item Is it the output language of formulating various kinds of prose structure that mimics the kinds we are used to people producing? 
\item Do people take naturally to the input because people are accustomed to discourse? 
\item Is the output valued because a story, right or wrong, is how people learn and remember? 
\item Is it the way it presents ideas with a persuasive stance that draws us in as any promoting stance does? 
\item Is it because the query can refine a previous result to make it even more appropriate?
\end {itemize}
All these factors help make GenAI attractive to users. But people using these tools also  note limitations.
The average ChatGPT user recognizes that the tool does not know the source of its knowledge, but tempers that with follow-up questions to guide the system to refine the answer. GenAI informs with narrative, but the structure can come off as formulaic and recognizable. They might still see these responses as a gift of something in the form they will need to produce, presented as an example they might even use. The scenario seems to speak to cognitive processes we are all familiar with from working with people. The results are presented in standard paragraphs and story structures we are used to. People recognize the results as well-produced first drafts. GenAI might lead users to more fully consider results and use search more powerfully to get to solutions. 
\paragraph{The Opportunities To Support People’s Explorations and Expositions Are Vast}

 Ideas come from a deep knowledge of many areas. The knowledgeable person, like repositories of online information, has representational and analytic knowledge of what they are working on. The “team” consists of people and AI-based knowledge resources that will have a range of solutions with purposeful idea-development discussions.

GenAI output responses being in specific coherent readable text softens incorrect responses with confident contextual style. With similar requests, today’s GenAI makes lists that repeat or recognize style and structure. 
 
Discursive acts of every kind are supported by GenAI. The GenAI revolution is evolving. Hundreds of plugins and Chrome extensions are available to work with ChatGPT. These apps add domain support and specialized capabilities. The list is impressive, supporting sophisticated mathematics, data visualization, education, presentation creation, productivity, program creation, and even video production. These all help with so many tasks we care about and value. But what helps us organize our browser tabs? A challenge is to integrate provenance and knowledge with its draft work products.

 \section{conclusion}
This paper explores when search engines or GenAI systems better serve users' knowledge needs
Our quantitative  purchasing experiment found even sceptical participants valued GenAI's efficiency for gathering knowledge to putting a decision in context.

Our qualitative comparison of 12 personas and knowledge gathering scenarios indicated that search engines can provide superior results for factual and niche knowledge. 

GenAI was more efficient and comprehensive for synthesizing perspectives in overviews or presentations. GenAI excels for questions requiring a verbal or media product to be created. 

In both experiments, the paradigms showed complementary strengths of the two knowledge access paradigm's suiting different goals. Advancing verifiable generative workflows could augment knowledge-building by combining provenance from search with contextual perspectives from AI as Perplexity attempts to do \cite{perplex}. While further research is needed, these findings suggest that integrating the unique advantages of search and GenAI could empower more robust knowledge exploration. Overall, this paper demonstrates the value each approach offers for varied user needs.

\subsection{Broader uses of LLM AI}
\hspace{1em} The last 20 years have shown the value of using keywords to access comprehensive knowledge sources. The value of these comprehensive data sources has changed the way we do everything, from play to work, to education, and to procurement. 

Search systems now take on many goals, from creating website like results to promoting goods and services, and even helping with physical directions. 

Until the current generation of GenAI, the work products we created with our knowledge seemed to be a different problem than finding the information. The value of a story to learn and remember is well known. GenAI also incorporates the context of recent requests in responses. Decisions are always made in contexts. While complex interfaces can be overwhelming, part of the user experience for GenAI might  become deploying reasonable search goals in context.  But real context has to be established by the writer; a GenAI story has to be put into one's own words to be yours and  believable.

The acquisition of facts has become fraught. Fake facts are found not just in GenAI’s hallucinations, but also in things people make up or repeat to support or create some strategic social or political movement. 
There used to be the idea that social media would help break down social barriers. It now seems that social media solving social problems was an early mirage. 
Fantasies are stories rooted enough in reality that people imagine and often hope they will become reality. More than hallucinations, GenAI creates fantasies. At least parts of fantasies are, or can become, reality. Without checking out what in fantasy is true or achievable, people can echo each other's fantasies in a spiral of confabulation. But communication and knowledge are important. 

The current opportunities for using AI to teach and provoke people to create more expansive and informed products are possibly the most important technical achievements of our time. It is up to us to make scenarios that improve the solutions people create. 

An important step towards all of this is assured truth and provenance, two features that search has honored and continue to be crucial for all real solutions. 
Provenance of where knowledge originated is critical to knowing what we know.  We look forward to a world in which peoples’ searches, work, and communication are well-informed, incisive, and factual. In another paper, we present an architecture called Generate and Search Test designed to help people assemble and evaluate alternative ideas \cite{selker2023ai}.  

We see a need for Knowledge Development Environments that as part of finalizing our work can help us compare and check different solutions to solve physical, human, social, societal, governmental, and environmental problems.

\section{comment to reviewers}
GenAI responses are changing and improving rapidly at the moment. We are willing fortify our experiment with more subjects using an updated ChatGPT if reviewers believe it is useful in the lead-up to publication.

\section{Acknowledgments}
\hspace{1em} This paper was made possible with the generosity of Google. 
I especially appreciate the exciting talks and encouragement of 
Scott Penberthy, 
the editing help of Ellen Shay, and Kathy Silke Prewitt, and the comments of others.

\newpage
\bibliographystyle{ACM-Reference-Format}
\scriptsize
\bibliography{chibib}

\newpage

 \section{Appendix}
 
\subsection{Appendix A: Data from Taxonomy of Knowledge Exploration Persona}

\begin {itemize} 
\item {\verb |Fact Checker|}
\begin {itemize}

\item {\verb |Search assumption:|} would provide the latest and most authoritative sources for the respective topics.
\item {\verb |ChatGPT assumption:|} might mention global databases like NASA's Climate Change and Global Warming data and resources, or refer to historical sources for World War II.
\item {\verb |Scenario 1:|} Sarah is a journalist preparing an article on climate change. She is looking for trustworthy data on global temperature changes. Her query might be: "What are the authoritative sources of global temperature data over the past 50 years?"
\item {\verb |Search|} shows articles from Climate.gov, NOAA's report, NASA, and Work meteorological organization. Users can directly read related articles.
\item {\verb |ChatGPT|} answers were more direct and focused 
\item {\verb |Scenario 2|} Bob is an author whose writing a historical novel. He needs to verify some facts about World War II. He might ask: "Can you provide sources on the number of casualties during World War II?"
\item {\verb |Search|} provides papers and articles which contain detailed graphics and data
\item {\verb |ChatGPT|} default model: can’t provide or cite sources in real-time, but gives an estimation.
\item {\verb |ChatGPT|} web browsing model(Beta) recommended the Wikipedia page
Scientist
Search This would provide the latest publications and active researchers in the respective fields.
\item {\verb |ChatGPT|} It may provide general information on who is considered a leader in the field and notable studies until its last update.
Scenario
\item {\verb |Scenario 1|}Dr. Johnson is researching gene therapy. He wants to stay updated on the latest findings. He might ask: "Who are the leading researchers in gene therapy and what are their recent publications?"
Search It offered more articles that allow use to pick what they want to read
\item {\verb |ChatGPT|} ChatGPT 4 web browsing model provided a few new projects of a top cell and gene therapy company. ChatGPT aggregated three articles and generated them for user to read.
\item {\verb |Scenario 2|} Lisa is studying astrophysics and needs to validate her thesis on black holes. She might ask: "What are the most recent studies on the formation of black holes?"
\item {\verb |Search VS. ChatGPT|} gave relative similar results because ChatGPT4 web browsing model grabbed the first articles of Bing
\end{itemize}
\item {\verb |Recreational Intellectual|}
\begin {itemize}
\item {\verb |Search assumption:|} Could provide detailed and up-to-date information on these topics.
\item {\verb |ChatGPT|} assumption: It might provide interesting facts about the solar system or Japanese customs based on the data it was trained on.
\item {\verb |Scenario 1|} Tom is a trivia enthusiast preparing for a quiz night. He might ask: "Can you provide me with interesting facts about the solar system?"
\item {\verb |Search VS. ChatGPT|} As a topic with a large amount of data, the knowledge provided by search and ChatGPT is similar As a topic with a larger amount of resources, the knowledge provided by Search and ChatGPT is similar
\item {\verb |Scenario 2|}Jane enjoys learning about different cultures. She might ask: "What are some unique customs and traditions in Japan?"
\item {\verb |Search|} The search article has a more comprehensive knowledge introduction on this subject, it presents a wide range of areas, such as greeting, entering houses, food, festivals, and arts.
\item {\verb |ChatGPT|} Had a more concise and focused direction
\end{itemize}
\item {\verb |Do It Yourself (DIY) Person|}
\begin{itemize}
\item {\verb |Search|} assumption:Could yield detailed guides, videos, or forums discussing these issues.
\item {\verb |ChatGPT| assumption:|}May provide a general step-by-step guide on building a garden shed or possible reasons why a car won't start.
\item {\verb |Scenario 1|} Anna wants to build a garden shed. She could ask: "How do I build a garden shed and what materials do I need?"
\item {\verb |Search|} The articles include steps, images, and videos, Search is much better than ChatGPT in terms of DIY. 
\item {\verb |ChatGPT|} Provide detailed steps, along with suggestions, such as general size, type recommendations, etc
\item {\verb |Scenario 2|} Peter's car won't start, he wants to try fixing it himself. He might ask: "What are the common reasons a car won't start?" 
\item {\verb |Search VS. ChatGPT|} gave similar answers. They provides common problems about the question.
\end {itemize}
\item {\verb |Hobbyist|}
\begin {itemize}
\item {\verb |Search assumption:|} It offers updated information, recent discussions, and newer platforms or locations
\item {\verb |ChatGPT assumption:|} Might suggest places to find stamp collectors online, or popular bird-watching locations in North America based on pre-existing knowledge.
\item {\verb |Scenario 1|} Richard is a stamp collector. He might ask: "Where can I find other stamp collectors to share my collection with?"
\item {\verb |Search|} shares more details about collectors and suggestions, and provides videos. In this scenario, Google search will be better
ChatGPT introduces multiple channels and parsing
\item {\verb |Scenario 2|} Emily loves bird-watching. She might ask: "What are the best locations for bird-watching in North America?"
\item {\verb |Search|} The problem triggered Google's traveling mode, which provides features such as plane prices, accommodations, maps, and more
\item {\verb |ChatGPT 4's|} web browsing model crawls the first article on the search page to provide 10 articles suitable for bird watching
\end {itemize}
\item {\verb |Merchant|}
\begin {itemize}
\item {\verb |Search|} assumption:offers the most recent trends, market analyses, and potential customer demographics.
\item {\verb |ChatGPT|} assumption: It could mention general trends or consumer behaviors based on data up to 2021.
\item {\verb |Scenario 1|} Lucy owns a boutique clothing store. She might ask: "What are the upcoming fashion trends in fall 2023?"
Search matched with the assumption. It provides the visualized answer. 
\item {\verb |ChatGPT 4|} web browning model gave some good recommendations. But it is not as good as Google search.
\item {\verb |Scenario 2|} Jack sells handmade wooden furniture. He could ask: "Who are my potential customers for handmade furniture and where can I find them?"
Search The search provides a glance at guidance, , frequently asked questions, as well as videos
\item {\verb |ChatGPT 4|} divides this problem into two parts. On the one hand, it provides professional categories, such as homeowner, interior designers, business, antique collector, etc., and then provides corresponding ways to find customers according to these categories. This way is much clearer and more compliant.
\end {itemize}
\item {\verb |Constructor|}
\begin {itemize}
\item {\verb |Search|} assumption:might offer the latest advancements, techniques, and best practices.
\item {\verb |ChatGPT|} assumption: Might provide general suggestions or methods for energy-efficient building or Python best practices based on its training.
\item {\verb |Scenario 1|} Susan is an architect planning to build an energy-efficient house. She might ask: "What are the latest methods and materials for building energy-efficient houses?"
Search displays articles that are more visual and easier for users to understand
\item {\verb |ChatGPT|} Web Browsing model takes the first article of search engine as reference and grabs the key points
\item {\verb |Scenario 2|} Robert is a software engineer building a new application. He might ask: "What are the best practices for coding in Python?"
\item {\verb |Search VS. ChatGPT|} showed similar results
\end{itemize}
\item {\verb |Professional|}
\begin {itemize}
\item {\verb |Search assumption:|} it provides the most recent job postings, available opportunities, and new courses or certifications.
\item {\verb |ChatGPT assumption:|} It could mention some popular job platforms and typical job opportunities for accountants or popular marketing certifications up to 2021.
Scenario
\item {\verb |Scenario 1|} Steve is an accountant considering a job change. He might ask: "What are the current job opportunities for accountants?"
Search can directly let the user click on the recruitment page, more in line with the user's requirements and habits.
\item {\verb |ChatGPT 4|} Web browsing model combated with two new articles and showed the newest information. The word "current" has triggered the web browsing behavior.
\item {\verb |Scenario 2 |} Linda is a marketing professional looking to enhance her skills. She could ask: "What top marketing courses or certifications should I consider?"
Search provides a number of recommended courses with advertizements, as well as blogs and lists of related questions frequently asked by users, which can bring users further reading
\item {\verb |ChatGPT|} provided more focused content, course names and introductions, and we can also ask for it to be provided with course links
\end{itemize}
\item {\verb |Consumer|}
\begin {itemize}
\item {\verb |Search assumption:|} This would offer the latest information on vacation trends or the most recent smartphone reviews.
\item {\verb |ChatGPT assumption:|} This might suggest some popular destinations or smartphones based on the training data up to 2021.
\item {\verb |Scenario1:|} Mary is planning a holiday and wants to explore options. She could ask: "What are some unique vacation destinations for summer 2023?"
\item {\verb |Search:|} For the recommendation of tourist attractions, Google directly displays pictures and hyperlinks, and after clicking, you can see the introduction of scenic spots directly in the search page, which is conducive to users to feel the scenery and characteristics of scenic spots.
\item {\verb |ChatGPT:|} The textual display of scenic spots does not bring much shock and temptation
\item {\verb |Scenario 2:|} John wants to buy a new phone. He might ask: "What are the top-rated smartphones in 2023?"
\item {\verb |Search Vs. ChatGPT:|} Default model of GPT4 cannot answer this question, and it advised what to look for in a top-rated smartphone, GPT4 Web Browsing model put the first article that it crawled as the reference, which was the same article from PC Magazine showing in Google search. It was very interesting.
Knowledgeable Domain Expert
\item {\verb |Search assumption:|} It can provide a range of resources such as online courses, video tutorials, scholarly articles, and textbooks that discuss sorting algorithms in-depth or offer comprehensive studies of Renaissance art.
\item {\verb |ChatGPT assumption: |}It can provide information on advanced chess strategies or the principles and practices of organic gardening based on its pre-existing knowledge.
\item {\verb |Scenario 1:|} Alex is an experienced chess player looking to improve his strategies. He could ask: "Can you provide advanced chess strategies that can enhance my game play?"
\item {\verb |Search Vs. ChatGPT:|} Google search recommends multiple videos on the home page to give users a more visual experience, while ChatGPT provides a comprehensive text parsing to facilitate user recording
\item {\verb |Scenario 2:|} Maria is a seasoned gardener, and she wants to extend her knowledge about organic gardening. She might ask: "What are the latest techniques in organic gardening?"
\item {\verb |Search Vs. ChatGPT:|} Because we designed the prompt to have the latest word, the Browsing Model in ChatGPT 4 triggers the search function, usually using the first article in the search result as a reference; ChatGPT 4 of the default model shows common organic gardening techniques
\end{itemize}
\item {\verb |Student|} 
\begin {itemize}
\item {\verb |Search assumption:|} It can provide a range of resources such as online courses, video tutorials, scholarly articles, and textbooks that discuss sorting algorithms in-depth or offer comprehensive studies of Renaissance art.
\item {\verb |ChatGPT assumption:|} It can explain sorting algorithms and their types, or analyze the artistic styles of the Renaissance period based on the data it was trained on.
\item {\verb |Scenario 1:|} David is studying computer science and is trying to understand algorithms better. He might ask: "Can you explain the concept of sorting algorithms and their types?"
\item {\verb |Search Vs. ChatGPT:|} Based on this topic, search and ChatGPT perform fairly well
\item {\verb |Scenario 2|} Emma is an art student learning about the Renaissance period. She might ask: "Can you provide an analysis of the artistic styles of the Renaissance period?"
\item {\verb |Search Vs. ChatGPT|} Match with the assumption
end{itemize}
\item {\verb |Communicator|}
begin{itemize}
\item {\verb |Search|} assumption:It can show examples of climate change blog posts or software manuals to provide inspiration or can suggest resources on how to improve writing skills, such as writing guides or courses.
\item {\verb |ChatGPT assumption:|} It can help to draft an introduction for a blog post on climate change impacts, or assist in creating an outline for a user-friendly software manual based on its knowledge of effective writing techniques and styles.
\item {\verb |Scenario 1:|} James is a blogger who's writing about climate change. He might ask: "Can you help me write an introduction to a blog post about climate change impacts?"
Search: Only provides good blogs as a reference, can not help users generate content
\item {\verb |ChatGPT:|} Automatically generate one or more articles for the user to select from, and users can even add more personalized style and points to the prompt
\item {\verb |Scenario 2:|} Rebecca is a technical writer preparing a user manual for new software. She might ask: "Can you help me outline a user-friendly manual for this software?"
\item {\verb |Search Vs. ChatGPT|} When the query relates to a regular product outline, Search and ChatGPT provide similar content, but if the user wants to add personal requirements, ChatGPT can generate a personalized Outline.
\item {\verb |Summary|}
This depends on the topic, if it is common ChatGPT gets the same answer. The issues is presentation style. Search gives so many choices and reasons to look at data. Unfortunately the range of search goals clutters the result focus. 
\end{itemize}
\end {itemize}

\subsubsection{Appendix B: Take-home messages}

Chat aggregated references making them easier to look through. 
Search gave broader results.
Search gave specific examples to explore .
Search gave better visualizations for spatial information.
Specialized websites have presentation advantages.
Web visual results again had advantages. Similar for common topics.
Search shows specific results for specific opportunities , ChatGPT prioritizes results without the bias of advertising for skill improvement goals.
Search is better for viewing results a lot of media or specific results are relevant.
Search was able to give videos. For common topics they were alike.
ChatGPT gave comprehensive results that helped. As a tutor Chat allowed followup questions that really help.
Chat responds in the form that the user is requesting content. 

\newpage
\subsection{ Appendix C: Formal experiment instructions - questionnaire}

\url{https://docs.google.com/forms/d/1_h46ddP90chUchY3l6aKV4r0N_AKTjZ4Qpdu4laHqp8/edit }

\begin{figure}
    \centering
    \includegraphics[width= 2.7in]{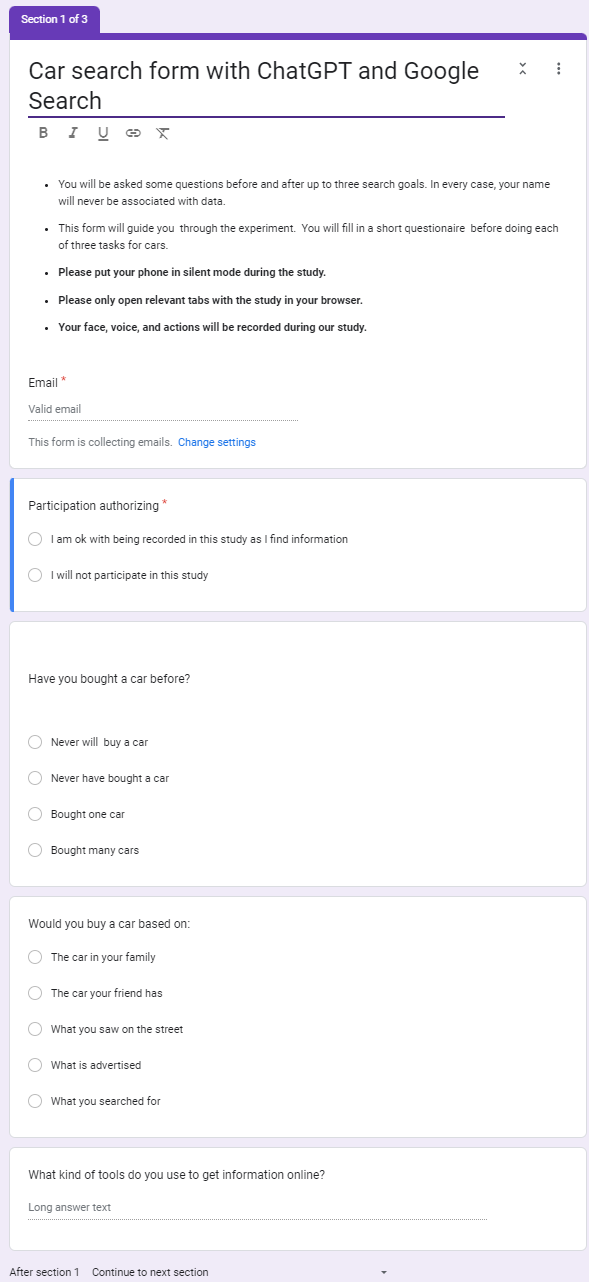}
    \caption{Online experiment instructions pretest}
    \label{Search/GenAI pretest}
    \Description{Participant questions: email, video authorization, have you bought a car before, would you buy a car based on, what information tools do you use online }
\end{figure}
\begin{figure}
    \centering
    \includegraphics[width= 2.7in]{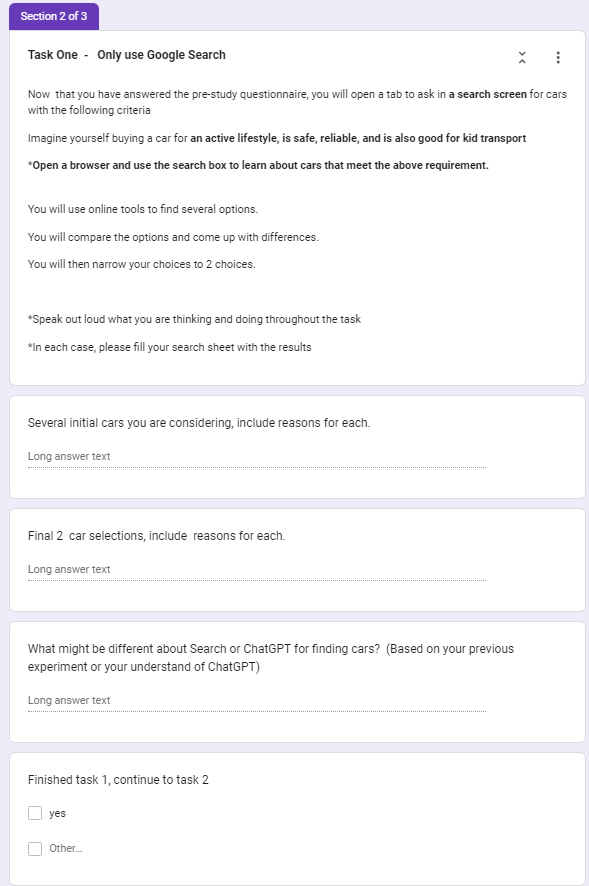}
    \caption{Online experiment task one instructions}
         \Description{Participant instructions: only use search , several cars you are considering,cars selection,  what might be different about Search of ChatGPT, finished task 1}
    \label{Search/GenAI task1 }
\end{figure}
\begin{figure}
    \centering
    \includegraphics[width= 2.7in]{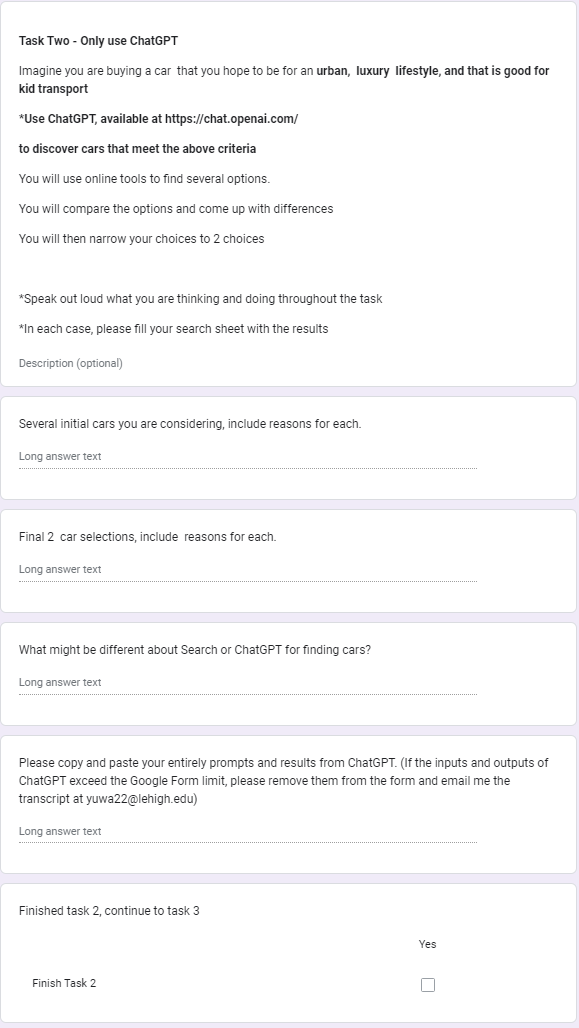}
    \caption{Online experiment task two instrutions}
    \Description{Participant instructions: Task2 use only ChatGPT, several initial cars, car selections, what my be different about search or ChatGPT, copy and paste prompts from Chat,}
    \label{Search GenAI task2}
\end{figure}
\begin{figure}
    \centering
    \includegraphics[width= 2.7in]{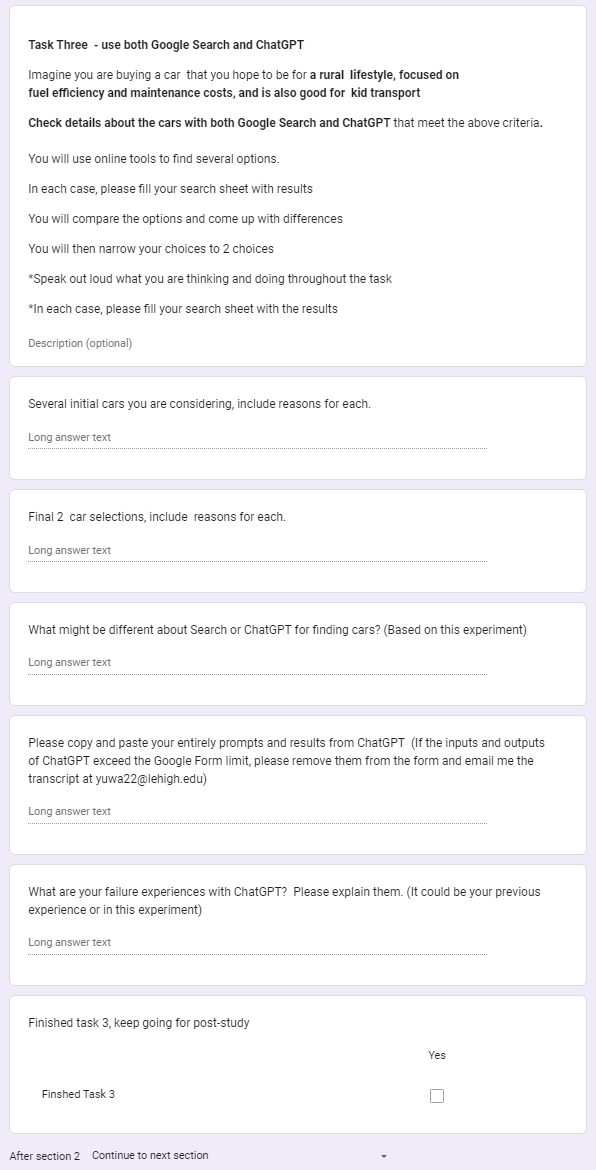}
    \caption{Online experiment task 3 instructions}
     \Description{Participant instructions: Task 3 , user both search and ChatGPT, several initial cars , final car selection, what might be different about Search and ChatGPT,  }
    \label{Search GenAI task 3}
\end{figure}
\begin{figure}
    \centering
    \includegraphics[width= 2.7in]{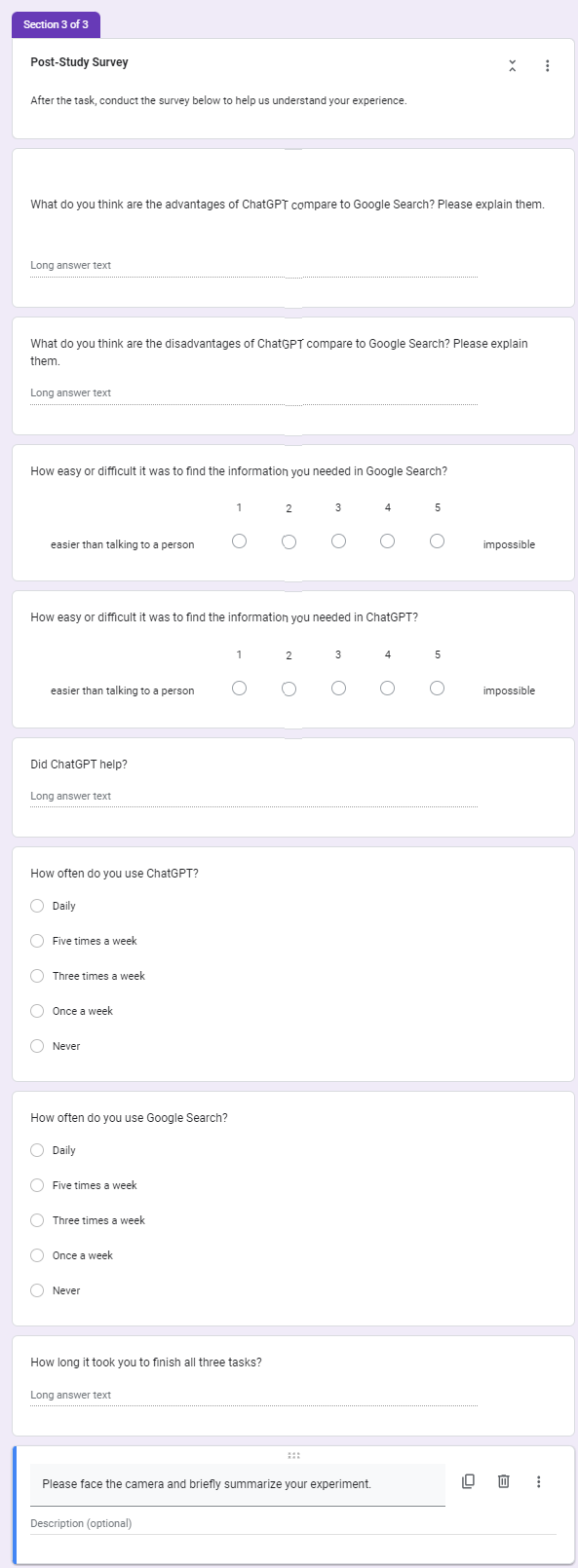}
    \caption{Online experiment Posttest}
    \label{searh GenAI posttest}
     \Description{Participant post-test: What do you think of the advantages of ChatGPT, What are the disadvantages, how easy was it to find information in Search,, how easy was it to find information in ChatGPT, How often do you use ChatGPT, How often do you use Search, How long did it take you to finish all three tasks?, summarize }
\end{figure}

\end{document}